\documentclass[11pt]{article}

\usepackage[a4paper]{geometry}
\usepackage[utf8]{inputenc}
\usepackage{graphicx,xcolor}
\usepackage{bm}
\usepackage{setspace}
\usepackage{amssymb}
\usepackage{amsmath}
\usepackage[round]{natbib}
\RequirePackage{lineno}

\newcommand{\eps}{\varepsilon}

\title{Effect of Creep on Corrosion-Induced Cracking}
\author{Ismail Aldellaa$^{1}$, Petr Havl\'{a}sek$^{2}$, Milan Jir\'{a}sek$^{2}$, Peter Grassl$^{1*}$}
\date{}
\begin{document}
\maketitle
\begin{center}
$^{1}$James Watt School of Engineering, University of Glasgow, Glasgow, UK\\
$^{2}$Czech Technical University in Prague, Czechia\\

  $^*$Corresponding author: Email: peter.grassl@glasgow.ac.uk\\

\end{center}

\section*{Abstract}
Corrosion-induced cracking is the most widely encountered and studied long-term deterioration process in reinforced concrete. Naturally occurring corrosion rates are so low that rust accumulates often over tens of years near the surface of the reinforcement bars before sufficient pressure in the surrounding concrete is generated to induce cracking in the concrete cover. To speed up the process in laboratory tests, corrosion setups with impressed currents have been developed in which the corrosion rate is controlled to be so high that cracking of the concrete cover occurs within a few days. Extrapolating the results of these accelerated tests to those of naturally occurring corrosion requires an understanding of the influence of long-term creep deformations of concrete on the corrosion-induced cracking process. In mathematical models in the literature, creep deformations are often ignored for accelerated but considered for natural corrosion rates in the form of an effective modulus.

In this work, three numerical models of increasing complexity are proposed with the aim to investigate the effect of creep on corrosion-induced cracking. The simplest approach is based on an uncracked axis-symmetric thick-walled cylinder combined with a plastic limit on the radial pressure-induced by the accumulation of rust. The model with intermediate complexity comprises a thick-walled cylinder model divided into an inner cracked and an outer uncracked layer. The most comprehensive model consists of a thick-walled cylinder discretised by a three-dimensional lattice approach. Basic creep is predicted in all three approaches by means of the B3 model developed by Ba\v{z}ant and co-workers. Time dependence of strength of concrete is modelled using \emph{fib} Model Code expressions.
It is shown that for the comprehensive lattice model, creep has limited influence on critical corrosion penetration, which indicates that the dependence of the critical corrosion penetration on corrosion rate must have other sources.

\section{Introduction}
The most commonly encountered deterioration process of steel reinforced concrete is corrosion-induced cracking \citep{bro97}.
Cracking occurs due to the formation of an expansive layer of corrosion products which is located close to the interface between concrete and reinforcement.
Predictive models for corrosion-induced cracking are here divided into three groups, namely uncracked axis-symmetric thick-walled cylinder models, cracked axis-symmetric thick-walled cylinder models and lattice models. All these models are mechanistic instead of imperical or semi-imperical.

The first two groups are based on the mechanics of an axis-symmetric thick-walled cylinder. These models are popular because the solution can be obtained quickly by solving an ordinary differential equation, without the need for spatial discretisation. Many articles have investigated the performance of these models, which are reviewed in \citet{JamAngAde13},~\citet{RosNoeMar20}~and~\citet{LiaWan20}.
Uncracked axis-symmetric thick-walled cylinder models construct a plastic limit on the radial pressure at the inner boundary. Examples of these uncracked cylinder models are \citet{Baz79b} and \citet{LiuWey98}. One of the well known shortcomings of this original approach is that the time to surface cracking is underestimated because the effect of material nonlinearities on the compliance of the cylinder are not taken into consideration. The overestimated stiffness leads to an underestimated displacement at the inner boundary at the time when the limit state is reached. This deficiency is alleviated by a newly proposed modification that takes into account the plastic strains and corrects the resulting compliance,  described in Section~\ref{sec:improved}. 

Cracked cylinder models still use the assumption of the axis-symmetric thick-walled cylinder, but divide the cylinder into an inner cracked and an outer uncracked layer, with the division progressing to the outer boundary during the corrosion process. The original idea of the two layers was proposed for bond between concrete and reinforcement in \citet{Tep79} and then adapted for corrosion-induced cracking in \citet{PanPap01}. Further modifications of these models were made in \citet{LiMelZhe06}, \citet{BhaGhoYas06} and \citet{CheValVol10}. These models have the advantage that they can describe the material nonlinearities and, therefore, predict the stiffness of the thick-walled cylinder correctly. The nonlinearity is introduced by transforming the opening of an assumed number of cracks into an axis-symmetric cracking strain. Once the cracked cylinder reaches the outer boundary, the assumption of axis-symmetry is not valid anymore and these thick-walled cylinder models cease to produce reliable results.

The lattice approaches are based on the spatial discretisation of the specimen using discrete elements within a matrix analysis approach. Many of the limitations of the thick-walled cylinder models are overcome by means of the spatial discretisation. No assumptions are required on the number of cracks. Furthermore, crack propagation beyond the occurrence of a surface crack can be modelled. However, approaches based on spatial discretisation are computationally much more time intensive and require techniques to model crack formation independent of the discretisation.
Many of these lattice approaches have been successfully used to model cracks in concrete at the meso-scale \citep{SchMie92b,MonCifMed17,ChaZhaSch20,ThiBadMen20} and macro-scale \citep{BolSai98,AsaAoyKim17}.
Lattice models can incorporate constitutive models, formulated in terms of tractions and displacement jumps, as commonly used in interface approaches for concrete fracture \citep{CabLopCar06,ZhoLu18}. An example of this is the damage plasticity lattice model used in \citet{GraDav11} and \citet{AthWheGra18}, which is used in this study. As shown in \citet{GraJir10}, these models yield element size-independent descriptions of crack openings  provided that fracture is localised. For initially distributed cracking such as for the case of corrosion-induced cracking, potential mesh dependence can be strongly reduced by introducing random fields of material strength as done by \citet{GraJir10}. This approach is adopted here as well. For the lattice approach used in this study, the spatial arrangement of the lattice elements and their cross-sectional properties are based on Delaunay and Voronoi tessellations of a set of random points placed in the domain as developed in \citet{YipMohBol05}. The random placement of nodes reduces the influence of the discretisation on the fracture patterns, as observed for other fracture approaches \citep{GraRem07,JirGra08}.

All the above approaches have in common that loading is applied by the formation of an expansive layer of corrosion products at the steel-concrete interface. These products are formed by corrosion, which is an electro-chemical process driven by the corrosion current density, normally so small that cracking is obtained only after many years \citep{bro97}. To accelerate the process in laboratory tests, an impressed current is applied, which increases the corrosion rate so that cracking can be obtained within days instead of years \citep{AndAloMol93}. To be able to use the results of these tests for predictive modelling of corrosion-induced cracking, it is required to relate the accelerated results to those of naturally occurring corrosion. Experimental studies
have investigated various aspects in which naturally occurring corrosion-induced cracking differs from accelerated tests using currents. For instance, corrosion products resulting from naturally occurring corrosion are not uniformly distributed around the steel bar circumference, but are more prominent in locations that are exposed to more severe conditions (e.g., direction of carbonation front and gradient of ingress of chlorides) \citep{ZhaKarWon11}. Furthermore, corrosion products forming in the vicinity of the steel-concrete interface are found to migrate into pores and cracks \citep{WonZhaKar10,MicPeaPet14,RobTenDij21}, which affects the build up of pressure and is often considered in modelling approaches \citep{LiuWey98,CheValVol10}. In addition, compaction of corrosion products has been investigated experimentally in \citet{OugBerFraFoc06} and incorporated in several modelling approaches \citep{Lun05a,BalBur11}. The composition of the corrosion products differs in naturally occurring corrosion from accelerated tests \citep{MaaSou03}. Therefore, it is expected that the influence of migration and compaction of corrosion products will differ in natural and accelerated corrosion tests.

Long-term effects of the surrounding concrete is another phenomenon which has not been studied extensively in the context of corrosion-induced cracking.
When concrete is subjected to sustained load, it exhibits creep deformations that grow in time.
Furthermore, strength and stiffness of concrete increase with age. This time-dependent response of concrete has been well studied in the literature \citep{HerMam65,Kommendant76,Bryant87,BazJir18}. However, its effect on corrosion-induced cracking has not been investigated in detail yet.
In the literature, corrosion-induced cracking models based on axis-symmetric thick-walled cylinder models often approximate the effect of creep for naturally occurring corrosion by replacing the short-term Young modulus with an effective modulus that is reduced by a creep coefficient. For accelerated corrosion, creep has often been disregarded, with the argument that cracking occurs after days instead of years. Very few experimental studies on the effect of creep on corrosion-induced surface cracking are available. In \citet{AloAndRod98}, a decrease of the corrosion rate (i.e., a decrease of the loading rate) has been shown to increase the crack opening, which was suggested to be related to creep. These results were obtained for the crack opening at the surface of the specimen, i.e., after surface cracking had already occurred. Similar results were also reported in more recent studies in \citet{PedAnd17}.

The aim of the present study is to investigate the influence of long-term effects on corrosion-induced surface cracking by numerical modelling. Three models of increasing complexity are used, namely uncracked and cracked axis-symmetric thick-walled cylinder models and lattice models.  Both the influence of basic creep and change of concrete maturity are studied. Migration of corrosion products into the adjacent concrete and compaction of the corrosion layer are not considered, so that the influence of long-term response of concrete can be isolated.
To achieve this aim, a new perfectly plastic model for fracture of a thick-walled cylinder is derived. Furthermore, a new formulation  for a cracked axis-symmetric thick-walled cylinder is developed. Finally, a new combination of a linear creep with a damage-plasticity constitutive model for a discrete approach is proposed.
To investigate the effect of creep, thick-walled cylinder analyses were carried out for corrosion rates corresponding to accelerated and natural corrosion. Furthermore, to better understand the influence of maturity, the analyses were carried out for corrosion starting at 28 and 10000 days, with the former value corresponding to accelerated experiments carried out in a research laboratory and the latter corresponding to naturally occurring corrosion of matured concrete.

\section{Modelling} \label{sec:model}

The effect of creep on corrosion-induced cracking of concrete is investigated by modelling a thick-walled cylinder in plane stress, which represents the top layer of a cylindrical specimen with a single reinforcement bar in the centre.
This setup is often used in accelerated corrosion tests.
This idealised geometry was chosen so that three models of increasing complexity can be applied, two of which rely on the assumption
of axial symmetry.
The three models are 
\begin{itemize}
    \item 
    an elastic axis-symmetric thick-walled cylinder model with a plastic limit, 
    \item 
    an inelastic axis-symmetric thick-walled cylinder model with advancing crack front, and 
    \item 
    a lattice model based on a three-dimensional discretisation of the thick-walled cylinder.
\end{itemize}
 Because of the different levels of complexity, it is not expected that these models will predict the same amount of corrosion at surface cracking. Therefore, it is not attempted to calibrate the models to produce the same results. Instead, input parameters are chosen based on previous modelling and experiments.

During the corrosion process, steel is transformed into corrosion products which, because of their lower density, produce the mechanical loading on the thick-walled cylinder. The amount of corrosion products is measured in the form of a corrosion penetration, i.e., the thickness of the steel layer that has been transformed into corrosion products. For small penetrations compared to the steel bar diameter, the corrosion penetration is proportional to the corrosion current density:
\begin{equation}\label{eq:basic1}
\Delta x_{\rm cor} = 0.0315\, i_{\rm cor} \Delta t
\end{equation}
Here, $\Delta x$ is the corrosion penetration in $\mu$m, $i_{\rm cor}$ is the corrosion current density in $\mu$A/cm$^2$, $\Delta t$ is the time duration in days, and $0.0315$ 
(in $10^{-4}$m$^3$/(A$\cdot$day))
is a conversion factor based on Faraday's law \citep{Bus00}. 

The focus of the present work is the effect of long-term behaviour of concrete, i.e., creep and change of maturity, on the initiation of surface cracking due to corrosion. To facilitate the assessment of the importance of the long-term behaviour, other effects such as migration and compaction of corrosion products, composition of rust products, shrinkage of concrete and external loading are not taken into account.
At surface cracking, the corrosion layer $\Delta x$ is so thin that it is acceptable to calculate the expansion as $\Delta u_{\mathrm{cor}} = (\alpha-1)\, \Delta x_{\mathrm{cor}}$ where $\alpha$ is an expansion factor which is set here to 2 as used in many studies before, \cite{MolAloAnd93}. Therefore, $\Delta u_{\mathrm{cor}} = \Delta x_{\mathrm{cor}}$.

The corrosion current density $i_{\rm cor}$ is the main variable in this study. Setting $i_{\mathrm{cor}} \geq 100$~$\mu$A/cm$^2$ provides conditions typically encountered in accelerated laboratory tests \citep{AndAloMol93}, in which surface cracking of the cylinder is reached in a few days. In some accelerated tests in \cite{MulSte09}, even higher current densities up to $1000$~$\mu$A/cm$^2$ were used. 
On the other hand, $i_{\rm cor} \leq 1$~$\mu$A/cm$^2$ corresponds to corrosion rates encountered in natural settings, where surface cracking is expected to be reached after several years \citep{bro97}.
The main equations of the three modelling approaches are described in the following subsections. Special focus is on the description of creep modelling. 

\subsection{Uncracked axis-symmetric model with plastic limit}

\subsubsection{Original formulation}

The first modelling approach is based on a combination of the elastic response of a thick-walled cylinder in plane stress combined with a plastic limit. It is commonly used as a mechanics based approach for predicting corrosion-induced cracking \citep{Baz79a,Baz79b,LiuWey98}. The main equations of the elastic model used here are reviewed and it is explained how creep and aging of concrete are considered.

\begin{figure}
  \begin{center}
    \begin{tabular}{cc}
      \includegraphics[width=7.5cm]{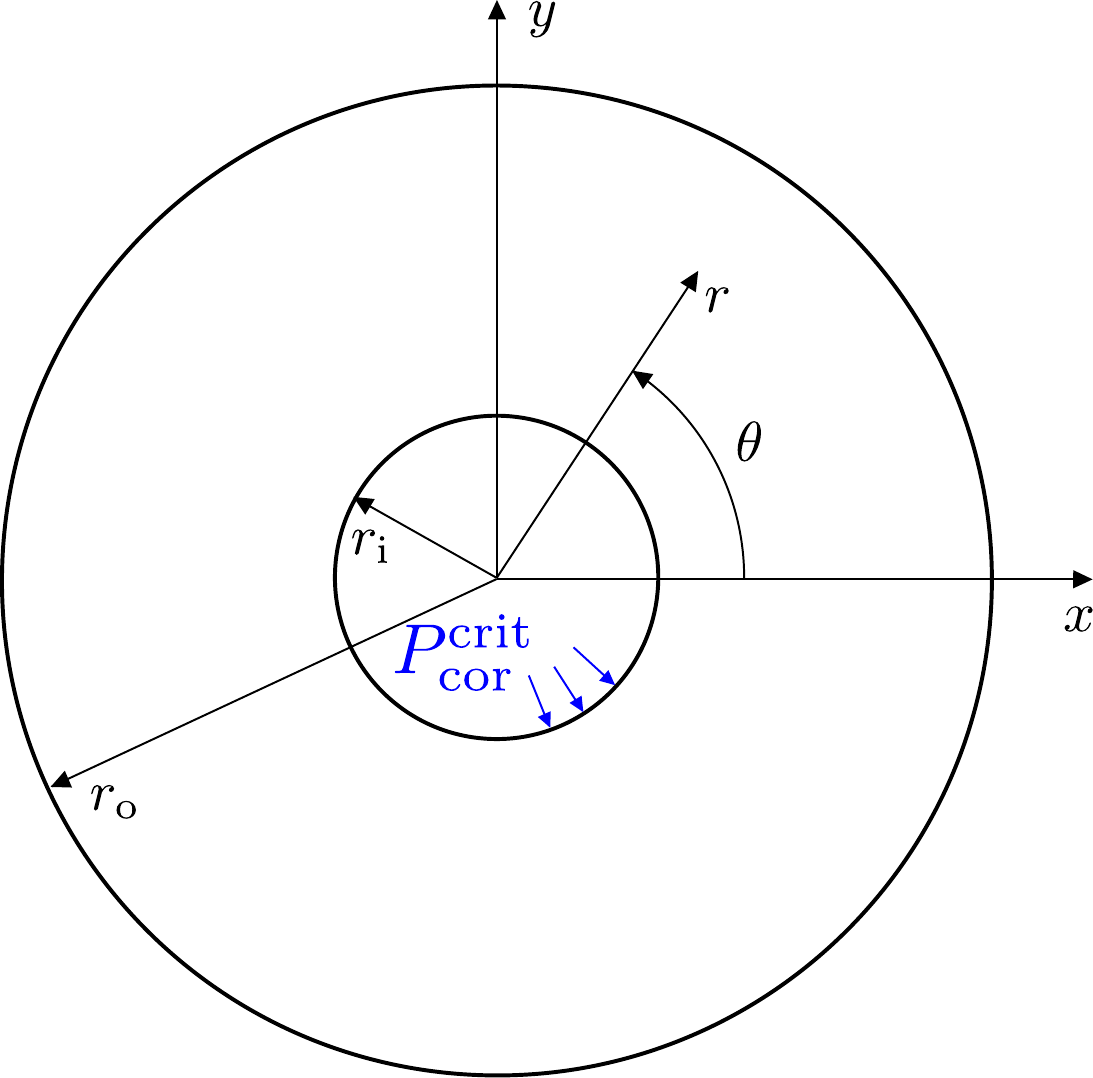} & \includegraphics[width=5.5cm]{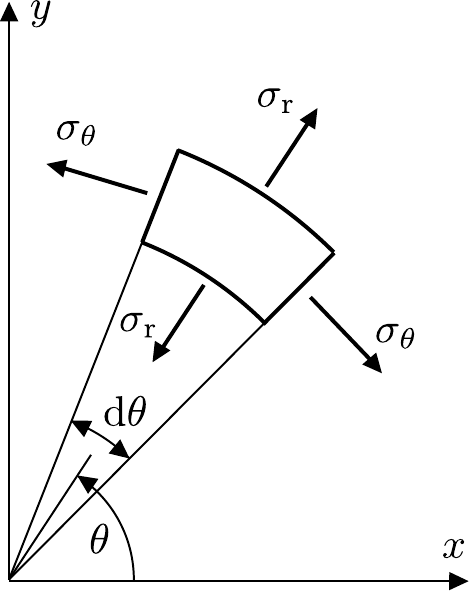}\\
      (a) & (b)
    \end{tabular}
  \end{center}
  \caption{Uncracked cylinder with plastic limit: a) Geometry of the thick-walled cylinder in polar coordinate system and b) small member in equilibrium.}
  \label{fig:elasticGeometry}
\end{figure}

For plane stress conditions, the constitutive equations linking strain and stress are
\begin{equation}\label{eq:elastic1}
\begin{Bmatrix}
  \eps_{\rm r} \\
  \eps_{\rm \theta}
\end{Bmatrix} =
\dfrac{1}{E^{''}} 
\begin{pmatrix}
  1 & -\nu\\
  -\nu & 1
\end{pmatrix}
\begin{Bmatrix}
  \sigma_{\rm r} \\
  \sigma_{\rm \theta}
\end{Bmatrix}
\end{equation}
where $r$ and $\theta$ are subscripts referring to the radial and circumferential direction in the cylindrical coordinate system as shown in Figure~\ref{fig:elasticGeometry}.
Furthermore, $\nu$ is Poisson's ratio and
$E^{''}$ is the effective modulus, i.e., a modulus which takes into account the age at loading and the load duration and approximately replaces the viscoelastic operator.
The stresses $\sigma_{\rm r}$ and $\sigma_{\rm \theta}$ shown in Figure~\ref{fig:elasticGeometry}b
satisfy the equilibrium condition
\begin{equation} \label{eq:elastic2}
\dfrac{\mathrm{d} \sigma_{\rm r}}{\mathrm{d} r} r + \sigma_{\mathrm{r}} - \sigma_{\theta} = 0
\end{equation}
The strain components are linked to the radial displacement $u$ by the kinematic relations
\begin{equation} \label{eq:elastic2a}
\varepsilon_{\mathrm{r}} = \dfrac{\mathrm{d}u}{\mathrm{d}r} \hspace{0.5cm} \mathrm{and} \hspace{0.5cm} \varepsilon_{\theta} = \dfrac{u}{r}
\end{equation}
Solving for the stresses in \eqref{eq:elastic1}, setting them in to \eqref{eq:elastic2} and using kinematic relations in \eqref{eq:elastic2a}, the differential equation for the radial displacement $u$ results in
\begin{equation}\label{eq:elastic3}
\dfrac{\mathrm{d}^2 u}{\mathrm{d}r^2} + \dfrac{1}{r} \dfrac{\mathrm{d}u}{\mathrm{d}r} - \dfrac{u}{r^2} = 0
\end{equation}

This equation is combined with appropriate boundary conditions, which idealise the mechanical effect of the corrosion process.
We assume that $\sigma_{\rm r} = -P$ at $r=r_{\rm i}$  and  $\sigma_{\rm r} = 0$ at $r=r_{\rm o}$ (see Figure~\ref{fig:elasticGeometry}). Here, $P$ is the pressure at the inner boundary of the cylinder  induced by the formation of corrosion product (with compression considered as positive pressure). The solution of the boundary value problem reads
\begin{equation}
    u(r) = \frac{P}{E^{''}}\frac{\left((1-\nu)r^2+(1+\nu)r_{\rm o}^2\right)r_{\rm i}^2}{\left(r_{\rm o}^2-r_{\rm i}^2\right)r}
\end{equation}
The displacement $u_{\rm i}=u(r_{\rm i})$ induced at the inner boundary by pressure $P$ can be expressed as
\begin{equation}\label{eq:elastic4w}
u_{\rm i} = \frac{P}{E^{''}C}
\end{equation}
where
\begin{equation}
C = \dfrac{r_{\rm o}^2 - r_{\rm i}^2}{ \left((1-\nu)r_{\rm i}^2 + (1+\nu) r_{\rm o}^2\right)r_{\rm i}}
\end{equation}
is a geometry-dependent factor and the product 
$E^{''}C$ can be interpreted as the structural stiffness.

In the original version of the uncracked cylinder model, it is assumed that relation (\ref{eq:elastic4w}) is approximately valid also 
in the limit state, characterized by the critical pressure
$P^{\mathrm{crit}}_{\mathrm{cor}}$ and
the critical radial displacement $u^{\mathrm{crit}}_{\rm i}$. Therefore, we can write
\begin{equation}\label{eq:elastic4}
P^{\mathrm{crit}}_{\mathrm{cor}} = {E}^{''}C  u^{\mathrm{crit}}_{\rm i}
\end{equation}
By introducing a plastic limit on the circumferential stress equal to the tensile strength ${f}_{\mathrm{t}}$, we determine the critical  pressure 
\begin{equation}\label{eq:elastic5}
P_{\rm cor}^{\mathrm{crit}} = {f}_{\mathrm{t}} \dfrac{r_{\rm o} - r_{\rm i}}{r_{\rm i}}
\end{equation}
and combining this with \eqref{eq:elastic5} we get
\begin{equation} \label{eq:elastic6}
u_{\rm i}^{\mathrm{crit}} = \dfrac{f_{\rm t}}{{E}^{''}C} \dfrac{r_{\rm o} - r_{\rm i}}{r_{\rm i}}
\end{equation}
The corrosion penetration at cracking is easily determined as $\Delta x_{\rm cor}^{\mathrm{crit}}=u_{\mathrm{i}}^{\mathrm{crit}}/(\alpha-1)$.
Finally, using the link between time and the penetration given by \eqref{eq:basic1}, we can
evaluate the time at which the plastic limit is reached,
\begin{equation} \label{eq:elastic7}
\Delta t^{\mathrm{crit}} =  \dfrac{u_{\mathrm{i}}^{\mathrm{crit}}}{0.0315 \left(\alpha - 1 \right) i_{\mathrm{cor}}} =  \dfrac{f_{\rm t}}{0.0315 \left(\alpha - 1 \right) i_{\mathrm{cor}} {E}^{''}C} \dfrac{r_{\mathrm{o}}-r_{\mathrm{i}}}{r_{\mathrm{i}}}
\end{equation}

In \eqref{eq:elastic7}, the effective modulus ${E}^{''}$ and the tensile strength $f_{\rm t}$ are time-dependent. The effective modulus is determined using the modified Age-Adjusted Effective Modulus Method \citep{Baz72,BazJir18}. The original formulation of the AAEM method in \citet{Baz72} exhibits a singularity at $\Delta t = 0.01$~days, which is the loading duration for which the short term effective modulus is determined. This is not a problem if the load duration is much larger than 0.01~days.
In the present study, we aim to investigate the effect of the corrosion density current including values which result in very short loading. To avoid any problems caused by the singularity, a modified expression for AAEM described in Example 4.5 in \citet{BazJir18} is used, which provides ${E}^{''} = {E}^{28}$ for $\Delta t_{\rm s} = 0.01$~days, where ${E}^{28}$ is the Young modulus at 28 days for a loading duration of $t = 0.01$~days. 
In this method, the effective modulus ${E}^{''}$ is given as
\begin{equation}\label{eq:elastic8}    
{E}^{''}\left(t, t_0\right) = \dfrac{1-R\left(t, t_0\right)J\left(t^*_0, t_0\right)}{J\left(t,t_0\right)-J\left(t^*_0,t_0\right)}
\end{equation}
where
\begin{equation}
  t_0^{*} = \left\{\begin{array}{ll}
  0.9t_0 + 0.1 t & \mbox{ if } t_0 < t < t_0 + 10 \Delta t_{\rm s}\\
  t_0+ \Delta t_{\rm s}  & \mbox{ if } t_0 + 10 \Delta t_{\rm s} \leq t
  \end{array}\right.
\end{equation}
and $t_0$ is the time at which loading starts.
In \eqref{eq:elastic8}, the relaxation function $R$ is approximated according to \citet{BazJir18} as
\begin{equation}\label{eq:elastic9}
  R\left(t, t_0\right) = \dfrac{1}{J\left(t,t_0\right)} \left[ 1 + \dfrac{c_1(t_0) J(t,t_0)}{10 J\left(t,t-\Delta t\right)} \left(\dfrac{J\left(t_m,t_0\right)}{J\left(t,t_{\rm{m}}\right)} - 1\right)\right]^{-10}
\end{equation}
where $\Delta t = 1$~day, $t_{\rm m} = (t+t_0)/2$, and
\begin{equation}
c_1\left(t_0\right) = 0.08 + 0.0119 \ln t_0
\end{equation}

The compliance function $J$ in \eqref{eq:elastic8} and \eqref{eq:elastic9} is chosen according to the B3 model originally proposed in \citet{Bazant94Baweja} and described in Section C.1 in \citet{BazJir18} as
\begin{equation} \label{eq:elastic10}
J\left(t, t_0\right) = q_1 + q_2 Q\left(t, t_0\right) + q_3 \ln\left[1+\left(t-t_0\right)^n\right] + q_4\ln\left(\dfrac{t}{t_0}\right)
\end{equation}
Here, $q_1$, $q_2$, $q_3$ and $q_4$ are creep parameters of the B3 model, $n=0.1$ and $Q$ is a special function defined by an integral formula, which
can be evaluated numerically or approximated analytically. Only basic creep is considered, i.e., shrinkage and drying creep are ignored. 

The dependence of the tensile strength on the age of concrete is taken into account using the \textit{fib} Model Code 2010.
Two expressions in this code are used. Firstly, formula
\begin{equation}\label{eq:elastic11}
{f}_c\left(t\right) = {f}^{28}_{\rm c} \exp\left(s\left[1-\sqrt{28/t}\right]\right)
\end{equation}
describes the increase in compressive strength with increasing age of the concrete. Here, ${f}^{28}_{\rm c}$ is the compressive strength at 28 days, $s$ is a parameter that depends on the aggregate type and $t$ is the concrete age substituted in days.
Secondly, the Model Code 2010 expression which relates tensile and compressive strengths is used to calculate the tensile strength at time $t$ and 28~days using ${f}_c\left(t\right)$ and ${f}^{28}_{\rm c}$, respectively.
The ratio of these two tensile strength values is then multiplied with the reference tensile strength to obtain the tensile strength at time $t$.
Reduction of the tensile strength due to nonlinear creep was not taken into account.

Setting \eqref{eq:elastic8} and \eqref{eq:elastic11} in \eqref{eq:elastic7}, a nonlinear equation for $t^{\mathrm{crit}}$ is obtained. An interactive method is then used to solve for $t^{\mathrm{crit}}$ and then to determine $x_{\mathrm{cor}}^{\mathrm{crit}}$ using \eqref{eq:basic1}. The calibration of the model is discussed in Section~\ref{sec:analyses}.

\subsubsection{Improved formulation}\label{sec:improved}

The simple version of the thick cylinder model used in the literature and presented in the previous subsection is based on
a somewhat inconsistent set of assumptions: Relation (\ref{eq:elastic4}) that
links the radial pressure on the internal boundary to the radial displacement is derived assuming linear (visco)elastic behavior but the limit value of the pressure (\ref{eq:elastic5})
is obtained from equilibrium in the plastic limit state, in which the circumferential stress in the
whole cylinder is at the yield limit, $f_{\rm t}$. 
Of course, to reach this stress distribution, 
the plastic zone must spread from the inner boundary
to the whole cylinder and the developed nonuniform plastic strains contribute to the overall deformation
of the cylinder and thus also to the boundary displacement. Under certain reasonable assumptions,
the stresses, strains and displacements in the plastic
limit state can be evaluated analytically,
as will be shown next.

The assumption that the circumferential stress
in the plastic limit state is uniform and equal to
the tensile yield stress naturally corresponds to
the Rankine yield condition, based on the maximum
principle stress. When the expected circumferential stress distribution
$\sigma_{\theta}(r)=f_{\rm t}$ is substituted into the equilibrium
equation (\ref{eq:elastic2}), one gets a differential equation
\begin{equation} \label{eq:elastic2x}
\dfrac{\mathrm{d} \sigma_{\rm r}}{\mathrm{d} r} r + \sigma_{\mathrm{r}} = f_{\rm t}
\end{equation}
from which the radial stress can be obtained.
Since the left-hand side of (\ref{eq:elastic2x})
corresponds to the derivative $\mathrm{d}(r\sigma_{\rm r})/\mathrm{d}r$,
integration is easy and leads to 
\begin{equation} 
\sigma_{\rm r}= f_{\rm t} + \frac{C_1}{r}
\end{equation}
When the integration constant $C_1=-r_{\rm o}f_{\rm t}$ is determined from the boundary condition $\sigma_{\rm r}(r_{\rm o})=0$, it turns out that the radial stress distribution in the plastic limit state is described by
\begin{equation}\label{eq:sigmar} 
\sigma_{\rm r}= f_{\rm t}\left(1-\frac{r_{\rm o}}{r}\right)
\end{equation}
The second boundary condition, $\sigma_{\rm r}(r_{\rm i})=-P_{\rm i}$, can be used to evaluate the radial pressure $P$ on the internal boundary in the plastic limit state. Interestingly, the result   exactly corresponds to the pressure  $P_{\rm cor}^{\rm crit}$
given by formula (\ref{eq:elastic5}). This is of course due to the
fact that the global equilibrium condition from which
(\ref{eq:elastic5}) was derived can be constructed by
integrating equation (\ref{eq:elastic2x}) from $r_{\rm i}$ to $r_{\rm o}$ and substituting 
$\sigma_{\rm r}(r_{\rm i})=-P_{\rm cor}^{\rm crit}$ and $\sigma_{\rm r}(r_{\rm o})=0$.

The next step is to proceed from stresses to strains.
For the plastic part of the model based on the Rankine
yield condition, it is natural to use the associated flow rule. Note that the radial stress  
$\sigma_{\rm r}$ given by (\ref{eq:sigmar}) is negative inside
the whole interval $(r_{\rm i},r_{\rm o})$.
This confirms the assumption that $\sigma_{\theta}=f_{\rm t}$ is the maximum principal stress,
and the flow rule associated with the Rankine condition predicts plastic flow with $\eps_{\theta {\rm p}}>0$ and $\eps_{\rm rp}=0$.
The total strains are the sum of these plastic strain components and the elastic strain components evaluated from the already known stresses using 
(\ref{eq:elastic1}). The resulting expressions are
\begin{eqnarray}\label{eq:epsr}
\eps_{\rm r} &=& \frac{f_{\rm t}}{E^{''}}\left(1-\nu-\frac{r_{\rm o}}{r}\right)\\
\label{eq:epst}
\eps_{\theta} &=& \frac{f_{\rm t}}{E^{''}}\left(1-\nu+\frac{\nu r_{\rm o}}{r}\right)+\eps_{\theta {\rm p}}
\end{eqnarray}
where the plastic strain $\eps_{\theta {\rm p}}$ is yet to be determined. 

The strain components are both linked to one displacement function, $u$, and so they must satisfy 
a certain compatibility condition. 
From the kinematic equations (\ref{eq:elastic2a}) we get ${\rm d}u/{\rm d}r=\eps_{\rm r}$ and $u=r\eps_{\theta}$, which can be simultaneously true
only if 
\begin{equation}
  \eps_{\rm r} = \frac{\rm d}{{\rm d}r}\left(r\eps_{\theta}\right)  
\end{equation}
Substituting from (\ref{eq:epsr})--(\ref{eq:epst})
and rearranging,
we obtain a differential equation
\begin{equation}
  \frac{\rm d}{{\rm d}r}\left(r\eps_{\theta {\rm p}}\right)  =  - \frac{f_{\rm t}}{E^{''}} \frac{r_{\rm o}}{r} 
\end{equation}
from which the plastic strain can be evaluated.
The general solution reads
\begin{equation}
  \eps_{\theta {\rm p}}  =  - \frac{f_{\rm t} r_{\rm o}}{E^{''}} \frac{\ln r}{r} + \frac{C_2}{r}
\end{equation}
The integration constant $C_2$ needs to be determined
from a suitable boundary condition. We are interested
in the state at the onset of plastic collapse,
when the complete yield mechanism has just developed. Therefore, there must be a point at which the stress
has just reached the yield limit and the plastic strain is still zero while all other points are at
a positive plastic strain. The plastic zone is expected to progress from the inner surface, and thus it can be expected that the point that yields last
is located at the outer surface. Condition
$\eps_{\theta {\rm p}}(r_{\rm o})=0$ leads to
$C_2 = (f_{\rm t}/E^{''})\, r_{\rm o}\ln r_{\rm o}$
and 
\begin{equation}
  \eps_{\theta {\rm p}}  =   \frac{f_{\rm t}}{E^{''}}
  \frac{r_{\rm o}}{r}\ln\frac{r_{\rm o}}{r} 
\end{equation}
Substituting this back into (\ref{eq:epst}) and using the kinematic equation, we can evaluate the displacement function 
\begin{equation}
    u = r\eps_{\theta} = \frac{f_{\rm t}}{E^{''}}\left((1-\nu)r+\nu r_{\rm o}\right)+r\eps_{\theta {\rm p}}
    =
    \frac{f_{\rm t}}{E^{''}}\left((1-\nu)r+\left(\nu+\ln\frac{r_{\rm o}}{r}\right)r_{\rm o}\right)
\end{equation}
Specifically, the displacement at the inner boundary
in the plastic limit state is given by
\begin{equation}
    u_{\rm i}^{\rm plast} =  
    \frac{f_{\rm t}}{E^{''}}\left((1-\nu)r_{\rm i}+\left(\nu+\ln\frac{r_{\rm o}}{r_{\rm i}}\right)r_{\rm o}\right)
\end{equation}
This consistently evaluated displacement can be related to the pressure in the plastic limit state by a formula of the same kind
as (\ref{eq:elastic4}), just with a modified geometry factor
\begin{equation}
    C^{\rm plast} = \frac{P_{\rm cor}^{\rm crit}}{E^{''}u_{\rm i}^{\rm plast}} =
    \frac{r_{\rm o}-r_{\rm i}}{\left((1-\nu)r_{\rm i}+\left(\nu+\ln\dfrac{r_{\rm o}}{r_{\rm i}}\right)r_{\rm o}\right)r_{\rm i}}
\end{equation}

Same as for the original model, this modified factor 
is used in (\ref{eq:elastic7}) considered as a nonlinear equation
from which the time to cracking can be evaluated.
The modified formula gives a lower stiffness
than the original formula
because the contribution of plastic yielding to the
overall compliance is taken into account.
It can therefore be expected that the resulting
times to cracking will be longer.

\subsection{Cracked axis-symmetric model}

The next formulation used in this study is for a cracked axis-symmetric thick-walled cylinder. In this model, the cylinder is split into a cracked and uncracked part. The limit on the inner pressure is obtained automatically by increasing incrementally the radius of the cracked cylinder. The way how cracking is introduced is conceptually based on previous work in \citet{GraJirGal19} in which the initiation of fluid-induced fracture of a thick-walled hollow permeable sphere was modelled. Here, it is presented for the first time for an impermeable thick-walled axis-symmetric cylinder.   
The amount of corrosion penetration required to induce surface cracking is strongly dependent on the stiffness of the surrounding material. Therefore, considering the cracking process within the cylinder is important.
Within an axis-symmetric thick-walled cylinder formulation, cracking is considered by introducing a cracking strain.
Many of the equations are the same as in the previous elastic modelling approach.
Here, the main differences are pointed out. 

\begin{figure}
  \begin{center}
    \begin{tabular}{cc}
      \includegraphics[width=6cm]{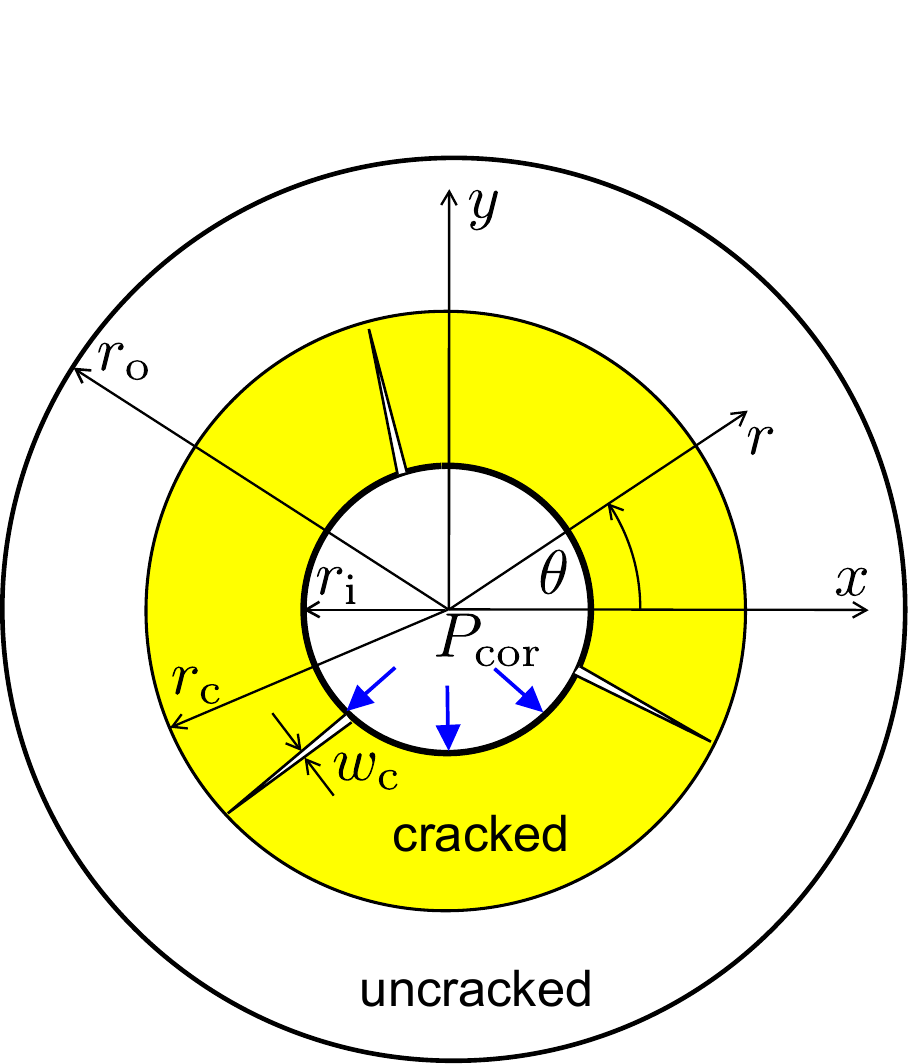} & \includegraphics[width=7.cm]{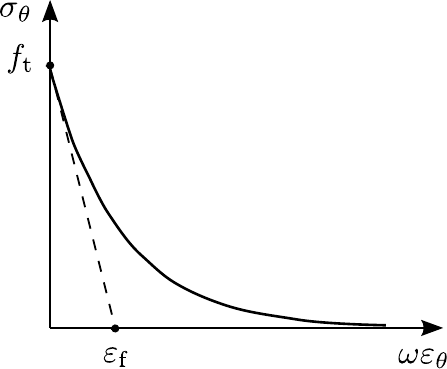} \\
      (a) & (b)
    \end{tabular}
  \end{center}
  \caption{Cracked cylinder: a) Geometry of the thick-walled cylinder split into a cracked and uncracked part and b) stress-cracking strain softening law.}
  \label{fig:crackedGeometry}
\end{figure}

The geometry of the cracked thick-walled cylinder is shown in Figure~\ref{fig:crackedGeometry}a.
The stress-strain equations 
\begin{equation}\label{eq:cracked1}
\begin{Bmatrix}
  \eps_{\rm r} \\
  \eps_{\rm \theta}
\end{Bmatrix} =
\dfrac{1}{{E}^{''}} 
\begin{pmatrix}
  1 & -\nu\\
  -\nu & 1
\end{pmatrix}
\begin{Bmatrix}
  \sigma_{\rm r} \\
  \sigma_{\rm \theta}
\end{Bmatrix}
+
\begin{Bmatrix}
  0 \\
  \varepsilon_{\rm \theta}^{\rm cr}
\end{Bmatrix}
\end{equation}
are an extension of the elastic law (\ref{eq:elastic1}) and incorporate the cracking strain $\eps_\theta^{\rm cr}$.
Before crack initiation, the cracking strain is zero. The crack is initiated when the
circumferential stress attains the tensile strength, $f_{\rm t}$. After crack initiation,
the cracking strain $\eps_\theta^{\rm cr}$ is linked to the circumferential stress $\sigma_\theta$
by the exponential softening law (Figure~\ref{fig:crackedGeometry}b) as
\begin{equation}\label{eq:cracked3}
\sigma_\theta = f\left(\eps_\theta^{\rm cr},r\right) \equiv  f_{\rm t}\exp\left(-\dfrac{\eps_\theta^{\rm cr}}{\varepsilon_{\rm f}}\right)   
\end{equation}
where $\varepsilon_{\rm f}= n_{\rm c} G_{\rm f}/(f_{\rm t} 2 \pi r)$ is a characteristic cracking strain.
The number of radial cracks $n_{\rm c}$ is one of the input parameters of the model. 
The equilibrium and kinematic equations are the same as in \eqref{eq:elastic2} and \eqref{eq:elastic2a}, respectively.

Solving \eqref{eq:cracked1} for $\sigma_{\rm r}$ and $\sigma_{\rm{\theta}}$, setting the stresses into \eqref{eq:elastic2} and using the kinematic equations in \eqref{eq:elastic2a} gives
\begin{equation}\label{eq:cracked8} 
  \dfrac{\mathrm{d}^2 u}{\mathrm{d}r^2} + \dfrac{1}{r} \dfrac{\mathrm{d}u}{\mathrm{d}r} - \dfrac{1}{r^2} u  + \dfrac{1}{r} \left(1-\nu\right)\varepsilon_{\theta}^{\rm cr} - \nu \dfrac{\mathrm{d} \varepsilon_{\theta}^{\rm cr}}{\mathrm{d}r} = 0
\end{equation}
Here, the term $\mathrm{d} \varepsilon_{\theta}^{\mathrm{cr}}/\mathrm{d}r$ is solved by differentiating \eqref{eq:cracked3} with respect to $r$, which provides
\begin{equation}\label{eq:cracked9}
\dfrac{\mathrm{d} \varepsilon_{\theta}^{\rm cr}}{\mathrm{d}r} = \dfrac{1}{A} \left(\dfrac{\mathrm{d}u}{\mathrm{d}r} \dfrac{1}{r} - \dfrac{u}{r^2} + \nu \dfrac{\mathrm{d}^2u}{\mathrm{d}r^2}\right)
\end{equation}
Here, the factor $A$ is
\begin{equation}
A = 1 - \dfrac{f_{\rm t}}{E\varepsilon_{\rm f}}\left(1-\nu^2\right) \exp\left(-\dfrac{\varepsilon_{\theta}^{\mathrm{cr}}}{\varepsilon_{\rm f}}\right)
\end{equation}
Setting the expression for the derivative of the cracking strain in \eqref{eq:cracked9} into \eqref{eq:cracked8} gives the nonlinear ordinary differential equation of the form
\begin{equation} \label{eq:cracked10}
\dfrac{\mathrm{d}^2u}{\mathrm{d}r^2} + \dfrac{1}{r} \dfrac{\mathrm{d}u}{\mathrm{d}r} \dfrac{A-\nu}{A-\nu^2} - \dfrac{1}{r^2} u  \dfrac{A-\nu}{A-\nu^2}  + \dfrac{1}{r} \dfrac{A \left(1-\nu\right)}{A-\nu^2} \varepsilon_{\theta}^{\rm cr} = 0
\end{equation}
whereby the cracking strain $\varepsilon_{\theta}^{\rm cr}$ is defined implicitly by the nonlinear differential equation in \eqref{eq:cracked3}. The boundary conditions are the same as for the uncracked model. A constant radial displacement rate at the inner boundary is prescribed and the traction at the outer boundary is assumed to vanish.
The nonlinear equation in \eqref{eq:cracked10} with $\varepsilon_{\theta}^{\rm cr}$ in \eqref{eq:cracked3} is solved using MATLAB with the boundary value solver bvp4c.

\subsection{Lattice modelling}
This section describes the lattice modelling approach for corrosion-induced cracking.
The approach is a combination of previously developed lattice models  based on a damage-plasticity constitutive model, presented previously in \citet{GraDav11, AthWheGra18}, and the linear creep model proposed in \citet{JirHav14} based on micro--prestress solidification theory of concrete creep. The new contribution in the present work is the combination of these two modelling concepts.
This modelling approach is too complex to allow us to present all the equations in detail as it was done for the uncracked and cracked axis-symmetric models. Instead, the overall concept is introduced and only the combination of the damage-plasticity model and the linear creep model is presented in more detail.

\begin{figure}
    \begin{center}
    \begin{tabular}{cc}
      \includegraphics[width=8cm]{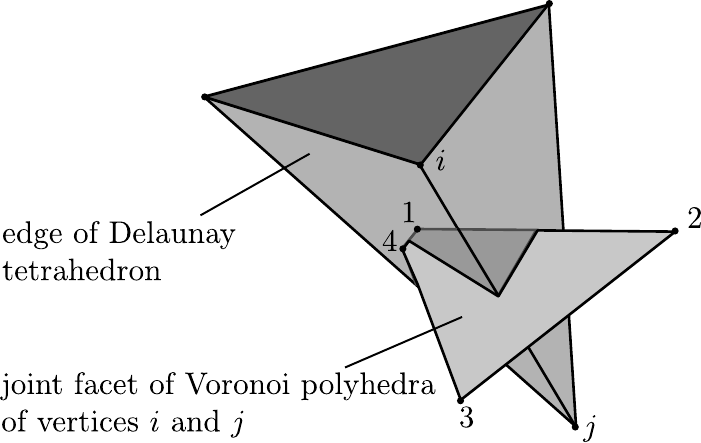} & \includegraphics[width=5.cm]{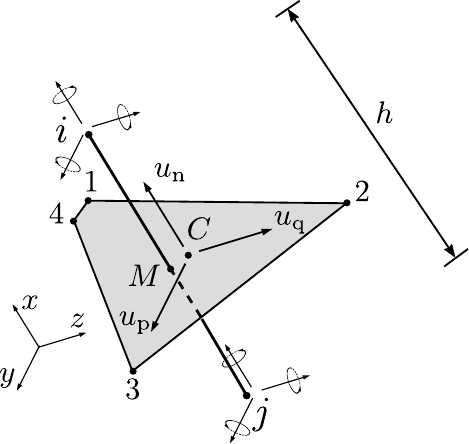}\\
      (a) & (b)\\
    \end{tabular}
    \begin{tabular}{cc}
    \includegraphics[width=10cm]{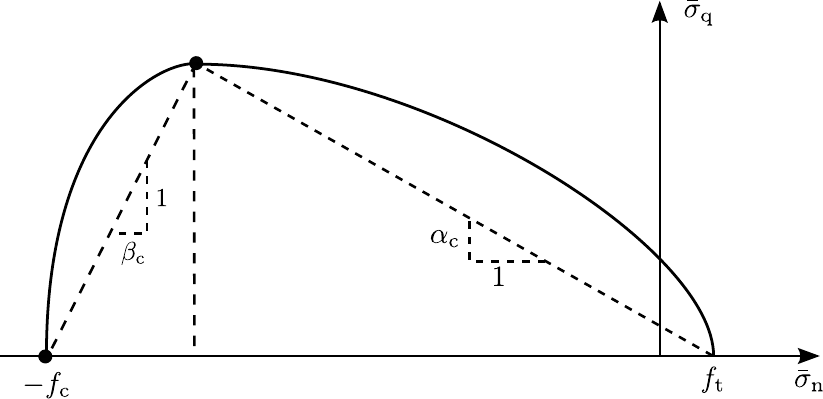} &  \includegraphics[width=5cm]{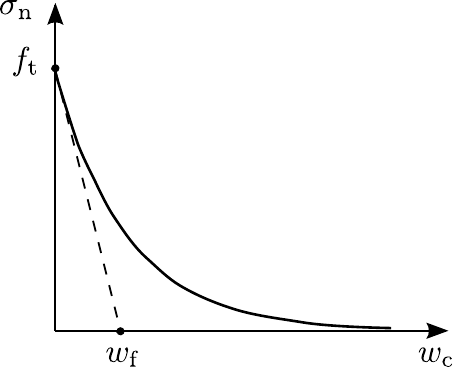}\\
    (c) & (d)
    \end{tabular}
    \end{center}
  \caption{Lattice model: a) geometrical relationship between Delaunay and Voronoi tessellations, b) lattice element with cross-section defined by the associated Voronoi facet, c) yield surface, and d) exponential softening law.}
  \label{fig:latticeOverview}
\end{figure}

The spatial discretisation is based on sequentially placed random vertices while enforcing a minimum distance. These vertices are used for dual Delaunay and Voronoi tessellations \citep{YipMohBol05}. The random vertices form the nodes of the lattice element (see vertices $i$ and $j$ in Figure~\ref{fig:latticeOverview}a). The edges of the Delaunay tetrahedra are used for the connections of the nodes, i.e. the edges give the location of the lattice elements. The mid-cross-section of the lattice elements is obtained from the facets of the Voronoi polyhedra. Boundaries of the specimen are modelled using mirrored vertices as described in \citet{YipMohBol05}. Each node has six degrees of freedom, namely three translations and three rotations. These nodal degrees of freedom are used to compute translational displacement jumps at the centroid of the mid-cross-section of the element by means of rigid body kinematics assuming that the two nodes of an element belong to two rigid polyhedra which meet at the facet which forms the mid-cross-section (Figure~\ref{fig:latticeOverview}b).
The translational displacement jump is transformed into strain by dividing them by the element length. This strain is used as an input for the constitutive model to compute the stress, which is then related to the nodal forces. No rotational stiffness at point $C$ is considered.
The constitutive model is described in more detail in the following section. 

\subsubsection{Damage plasticity model with extension to visco-elasticity}
For the damage plasticity model, the plasticity part uses a yield surface which is a function of the normal component and the norm of the two shear components of the effective stress. Geometrically, the surface is composed of two ellipses, so that there are limits on the normal component of the effective stress both in tension and in compression (Figure~\ref{fig:latticeOverview}c). Plastic flow is modelled with a plastic potential which differs from the yield function, so that the amount of normal plastic strain can be controlled. Damage is determined from the positive normal component of the plastic strain using an exponential softening law (Figure~\ref{fig:latticeOverview}d).
The damage-plasticity model uses nine parameters. Parameters $E$ and $a_1$ control the macroscopic Young's modulus and Poisson's ratio of the material. Additional parameters, $f_{\rm t}$, $f_{\rm c}$, $\alpha_c$, $\beta_c$ and $A_{\rm h}$, determine the shape and size of the yield surface and its evolution during hardening.
Furthermore, parameter $\psi$ controls the plastic flow. Finally, $G_{\rm F}$ controls the amount of energy dissipated during cracking.

For the standard damage plasticity approach, the stress-dependent strain vector is defined as
$\boldsymbol{\varepsilon}_{\sigma} = \boldsymbol{\varepsilon} - \boldsymbol{\varepsilon}_\mathrm{cor}$ where $\boldsymbol{\varepsilon}$ is the total strain and $\boldsymbol{\varepsilon}_{\mathrm{cor}}$ is the eigenstrain which reflects the formation of corrosion products.
Using first the plasticity part of the damage-plasticity model, the effective stresses $\bar{\boldsymbol{\sigma}}$, plastic strain $\boldsymbol{\varepsilon}_{\rm p}$ and hardening variable $\kappa_{\rm p}$ are computed. All these quantities are independent of damage. Subsequently, the damage parameter $\omega$ is determined using the results of the plasticity part. The nominal stress is determined as
\begin{equation}
  \boldsymbol{\sigma} = \left(1-\omega\right) \mathbf{D}_{\rm e} \left(\boldsymbol{\varepsilon}_{\sigma} - \boldsymbol{\varepsilon}_{\rm p}\right) = \mathbf{D}_{\rm e} \left(\boldsymbol{\varepsilon}_{\sigma} - \boldsymbol{\varepsilon}_{\rm p} - \omega \left(\boldsymbol{\varepsilon}_{\sigma} - \boldsymbol{\varepsilon}_{\rm p}\right)\right)
\end{equation}
Here, the inelastic strain which is subtracted from the stress-dependent strain $\boldsymbol{\varepsilon}_{\sigma}$ consists of an irreversible part in the form of the plastic strain $\boldsymbol{\varepsilon}_{\rm p}$ and a reversible part in the form of the damage strain
\begin{equation}
  \boldsymbol{\varepsilon}_{\omega} = \omega \left(\boldsymbol{\varepsilon}_{\sigma} - \boldsymbol{\varepsilon}_{\rm p}\right)
\end{equation}

For the visco-elastic extension of the damage-plasticity model, we consider the special case of constant ambient temperature under hygrally sealed conditions. For this case, the response of the visco-elastic model is identical to the basic creep compliance function of the B3 model. The implementation of the MPS material model uses the rate-type approach summarised in ``Algorithm 10.1: Incremental stress–strain relation according to the microprestress-solidification theory" \citep{BazJir18}. Under sealed conditions this algorithm can be further simplified to ``Algorithm 5.3: Exponential algorithm for solidifying Kelvin chain". Additionally, the parameters of the Dirichlet series are estimated from the continuous retardation spectrum \citep{JirHav14b} of the non-aging compliance function, which makes the implementation more efficient.

The visco-elastic extension of the damage-plasticity approach models the evolution of tensile and compressive strength with time described by the \textit{fib} Model Code 2010 formulae \citep{MC10}. The viscoelastic model and the damage-plasticity model are linked in series. This implies that the stress transmitted by the two models needs to be equal and the (stress-dependent) total strain $\boldsymbol{\varepsilon}_{\sigma}$ needs to be split into viscoelastic $\boldsymbol{\varepsilon}_\mathrm{ve}$ and inelastic components $\boldsymbol{\varepsilon}_\mathrm{pl}$ and $\boldsymbol{\varepsilon}_{\omega}$. Owing to the viscoelastic nature of the problem, the stiffness matrix is no longer constant and is proportional to the incremental modulus $E$. For simplicity, the stiffness which is used in the plasticity-damage part of the model is assumed as a constant and is evaluated from the basic creep compliance function using the expression for the conventional Young's modulus at the age of 28 days for a load duration of 0.01~days ($E_{28} = {1}/{J_b(28.01, 28)}$).
In the first step, the stress-dependent strain vector associated with the viscoelastic material is computed as
\begin{equation}
  \boldsymbol{\varepsilon}_{\mathrm{ve},\sigma} = \boldsymbol{\varepsilon} - \boldsymbol{\varepsilon}_{\mathrm{cor}} - \left(\boldsymbol{\varepsilon}_{\mathrm{p}} + \boldsymbol{\varepsilon}_{\omega}\right)
\end{equation}
where $\boldsymbol{\varepsilon}_\omega$ and $\boldsymbol{\varepsilon}_{\mathrm{p}}$ are the temporary values of damage and plastic strain vectors, respectively, which will be updated during the iterative process.   
The stress-dependent strain vector $\boldsymbol{\varepsilon}_{\mathrm{ve},\sigma}$ is then used in the stress-evaluation algorithm of the viscoelastic model and the stress $\boldsymbol{\sigma}_{\mathrm{ve}}(\boldsymbol{\varepsilon}_{\mathrm{ve},\sigma})$ is computed. This is the nominal stress, which needs to be equal to the nominal stress determined by the damage-plasticity model since the two models are combined in series.
Therefore, as the next step, the stress-dependent part of the strain vector associated with the plasticity-damage material is computed as
\begin{equation}
 \boldsymbol{\varepsilon}_{\sigma} = \mathbf{D}_{28}^{-1} \boldsymbol{\sigma}_{\mathrm{ve}} + \boldsymbol{\varepsilon}_\omega + \boldsymbol{\varepsilon}_{\mathrm{p}}
\end{equation}
where $\mathbf{D}_{28}$ is the stiffness matrix
that corresponds to the conventional elastic modulus,
$E_{28}$.
Strain $\boldsymbol{\varepsilon}_{\sigma}$ is then used for the damage-plasticity model following the steps outlined for the standard version of that model.
The nominal stress obtained from the damage-plasticity approach is compared to nominal stress obtained from the visco-elastic model. If the difference between these two nominal stresses exceeds a prescribed tolerance, the two nominal stresses are calculated again whereby the viscoelastic nominal stress is now determined with a new strain which considers the new plastic and damage parts of the inelastic strains.   
This procedure is repeated until the stresses obtained using the two models are the same.

\section{Analyses} \label{sec:analyses}

The three modelling approaches described in the previous section (uncracked and cracked axis-symmetric model, and lattice model) are calibrated using short-term material properties of a reference concrete at 28 days with Young's modulus ${E}^{\rm{ref}} = 30$~GPa, Poisson's ratio $\nu^{\rm{ref}}=0.2$, tensile strength $f_{\rm t}^{\rm{ref}} = 3$~MPa, compressive strength ${f}_{\rm c}^{\rm{ref}} = 30$~MPa and fracture energy ${G}_{\rm F}^{\rm{ref}} = 150$~N/m. The reference Young's modulus corresponds to an effective modulus with a load duration of 0.01~days at loading start time of $28$~days.
In the following paragraphs, the geometry and boundary conditions of the problem, the calibration procedure and the results obtained with the three approaches are presented.

\subsection{Geometry and boundary conditions}

\begin{figure}
  \begin{center}
    \includegraphics[width=6cm]{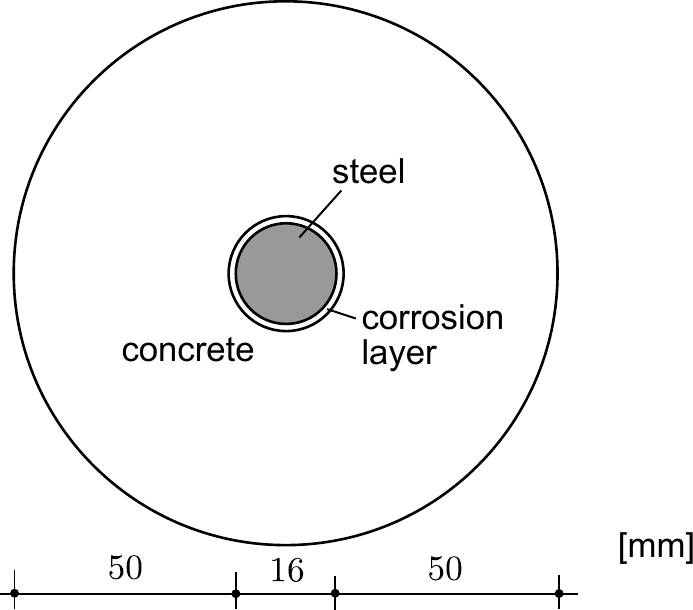}\\ 
  \end{center}
  \caption{Geometry of the thick-walled cylinder analysed with the three approaches. The out-of-plane thickness is $10$~mm.}
  \label{fig:geometry}  
\end{figure}

The geometry of the problem analysed consists of a concrete cylinder with a steel reinforcement bar at its centre.
The radius of the reinforcement bar is $r_{\rm i} = 8$~mm and the outer radius of the concrete cylinder is $r_{\rm o} = 58$~mm as shown in Figure~\ref{fig:geometry}. The out of plane thickness is $10$~mm.
This slice of thick-walled cylinder represents the region of a test specimen close to the free boundary. Therefore, plane stress conditions are assumed for the two axis-symmetric models. For the lattice model, the 3D slice was discretised and the nodes of the lattice were not restrained in the out-of-plane direction. For the two axis-symmetric models, the corrosion process is modelled by prescribing an outwards radial displacement at the inner boundary of the concrete cylinder. This assumes that the steel reinforcement bar is so stiff that its deformation does not influence the results. For the lattice model, it was more straight-forward to discretise the steel bar as well and model the radial displacement induced by the formation of the corrosion products by a layer of the lattice elements adjacent of the steel bar, whereby the elements are arranged so they are perpendicular to the circumference of the steel bar. Within this layer of lattice elements the expansion due to the formation of the corrosion products was modelled by an eigenstrain computed from a prescribed eigen-displacement, which was determined for all approaches using \eqref{eq:basic1} and the assumption that $u_{\rm cor} = x_{\rm cor}$, i.e.~$\alpha = 2$. This approach makes the radial displacement independent of the lattice element length. The main variable investigated is the corrosion rate $i_{\rm cor}$. Analyses with the elastic-axisymmetric and lattice models for different corrosion current densities $i_{\rm cor} = 0.1$, $1$, $10$, $100$, $1000$ and $10000$~$\mu$A/cm$^2$ are performed to determine the amount of corrosion penetration at which the maximum internal pressure, which is called strength of the thick-walled cylinder, is reached. For most analyses, surface cracking is reached at this critical corrosion penetration.
In general, the smaller the corrosion current density, the greater is the time at which the strength is reached. However, interplay of creep and change of maturity affects the time to cracking, too. To better understand the influence of maturity, the analyses were carried out for $t_0 = 28$~and~$10000$~days with the former corresponding to accelerated experiments carried out in a research laboratory and the latter corresponding to naturally occurring corrosion of matured concrete.

\subsection{Calibration}
The calibration procedures for the three models are described below.
For the uncracked thick-walled cylinder model with plastic limit, the model parameters are ${\nu}^{\rm{ref}}$, $f_{\rm t}^{\rm{ref}}$, $q_1$, $q_2$, $q_3$ and $q_4$. The first two parameters can be chosen directly from the properties of the reference concrete, which are ${\nu}^{\rm{ref}} =0.2$ and $f_{\rm t}^{\rm{ref}} = 3$~MPa. The four parameters for basic creep are determined in two steps. Firstly, an initial set of parameters are calculated using the formulas in \citet{BazBaw95}, assuming concrete mix composition as water $w = 180$~kg/m$^3$, cement $c = 360$~kg/m$^3$ and aggregate $a = 1860$~kg/m$^3$. These formulas estimate
the elastic modulus from the compressive strength
based on an empirical formula suggested by the ACI recommendations,
and so the resulting modulus differs from the Young modulus of the reference concrete at 28 days for a loading duration of 0.01~days. In the present study, both reference compressive strength and reference Young's modulus are given. To obtain the desired reference Young modulus ${E}^{\rm{ref}}$, the four parameters $q_1$ to $q_4$ are scaled by the same factor. The resulting creep parameters are $q_1 = 19.367\times10^{-6}$/MPa, $q_2 = 137.865\times10^{-6}$/MPa, $q_3 = 2.499\times10^{-6}$/MPa and $q_4 = 5.381\times10^{-6}$/MPa.
The cracked thick-walled cylinder model uses the same parameters as the uncracked cylinder model, but requires two additional parameters---the fracture energy ${G}_{\rm F}$ and the number of cracks $n_{\rm c}$. The fracture energy is set to ${G}^{\rm{ref}}_{\rm F} = 150$~N/m. The number of cracks needs to be assumed, which is one of the disadvantages of the cracked thick-walled cylinder model. In this study, we use $n_{\rm c} = 4$, which was used before in \citet{FahWheGal17} for a similar model covering the special case of Poisson's ratio $\nu = 0$.

The parameters of the lattice approach are calibrated in multiple steps. Firstly, the parameters $E$ and $a_1$ of the lattice model are determined from a direct tensile test so that the reference elastic properties ${E}^{\rm{ref}}=30$~GPa and $\nu^{\rm{ref}} = 0.2$ are obtained. These analyses were initially carried out without taking creep into account. The specimen used for the direct tensile test has a length of 75~mm, width of 50~mm and an out-of plane thickness of 10~mm. The load was applied in the long direction of the specimen. The values of the two parameters are determined as $E=45.91$~GPa and $a_1=0.297$. Next, the creep parameters of the B3 model $q_1$ to $q_4$ are determined on a single element subjected to uniaxial tension using the same approach as for the uncracked and cracked cylinder models, namely first determining the parameters using the formulas in \cite{BazBaw95} and then scaling them to obtain the lattice material Young's modulus $E$. This is the same approach that was used for the two axis-symmetric models, with the difference that the target Young's modulus for the lattice approach is $E=45.91$~GPa.
The resulting values of the creep parameter for the lattice model are $q_1 = 12.640\times10^{-6}$/MPa, $q_2 = 89.982\times10^{-6}$/MPa, $q_3 = 1.631\times10^{-6}$/MPa and $q_4 = 3.512\times10^{-6}$/MPa.
For the single element, the load is applied instantaneously at $t_0 = 28$~days and the secant modulus is determined at a duration of $\Delta t = 0.01$~days.
Next, the inelastic parameters are determined, so that $f_{\rm t}^{\rm{ref}}$ and ${G}_{\rm f}^{\rm{ref}}$ are obtained. This calibration is carried out on the direct tensile test which was used earlier for obtaining the elastic properties. To avoid fracture at the boundaries, the end regions were set to be elastic so that only a middle region of 50~mm was able to fracture.
The lattice model considers the randomness of the material strength by an autocorrelated random field with a Gaussian probability function of fully correlated strength and fracture energy \citep{GraJir10}. The autocorrelation length is chosen as $l_{\rm a} = 2.67$~mm. Furthermore, the coefficient of variation of the random field is $c_{\rm v} = 0.2$.
Six analyses with random fields and meshes were carried out for this calibration step. The input parameters of the lattice model $f_{\rm t}$ and $w_{\rm f}$ were chosen so that mean tensile strength and fracture energy determined from the six analyses agreed with the reference properties. The resulting input parameters are $f_{\rm t} = 2.35$~MPa and $w_{\rm f} = 0.02$~mm. The other parameters of this lattice model are set to their default values as described in the manual of OOFEM \citep{Pat12}. 

\subsection{Results}
\begin{figure}
  \begin{center}
    \includegraphics[width=12cm]{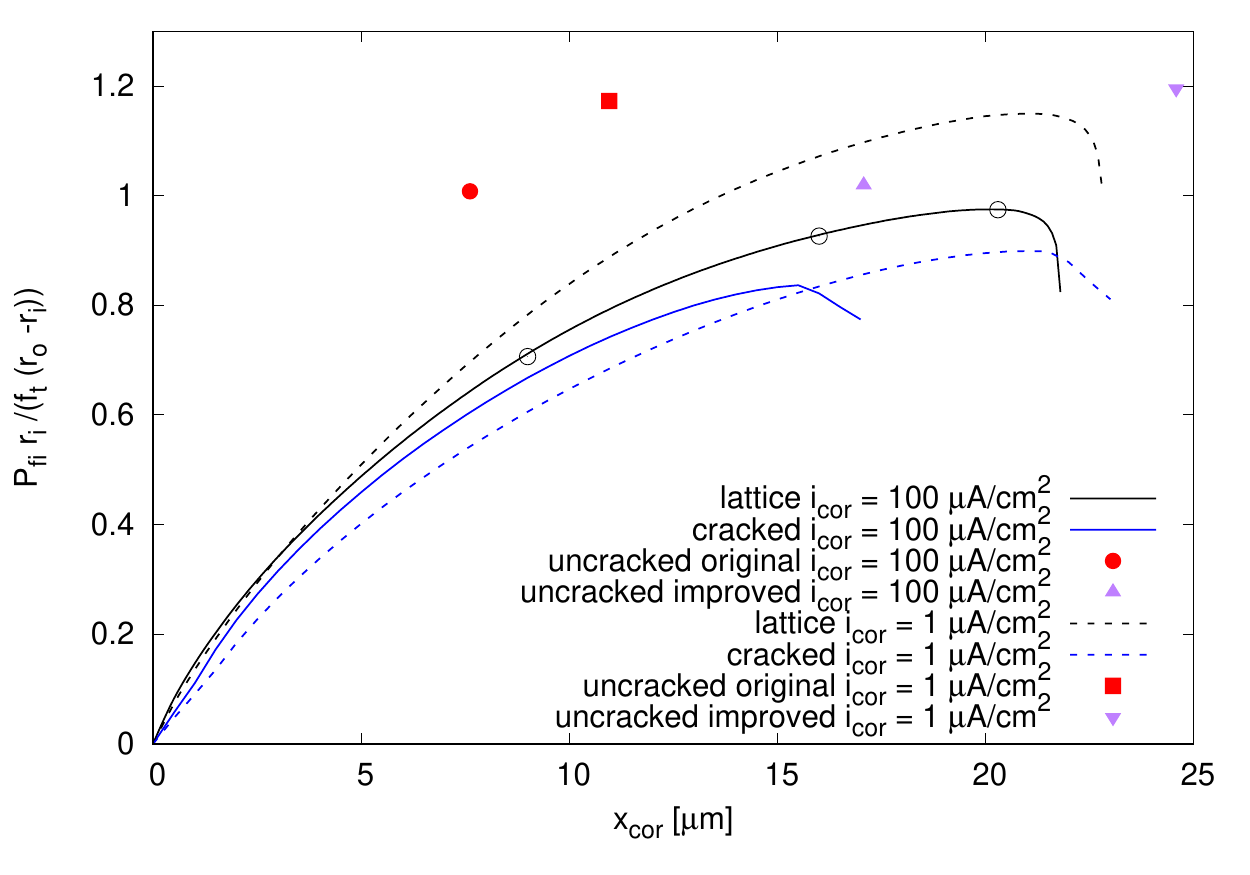}
  \end{center}
  \caption{Normalised pressure at inner boundary versus corrosion penetration $x_{\rm cor}$ for $t_0=28$~days. The symbols on the lattice curve mark stages at which crack patterns are shown in Figure~\ref{fig:latticeCylinderCrack}.}
  \label{fig:pressure28}
\end{figure}

\begin{figure}
  \begin{center}
    \includegraphics[width=12cm]{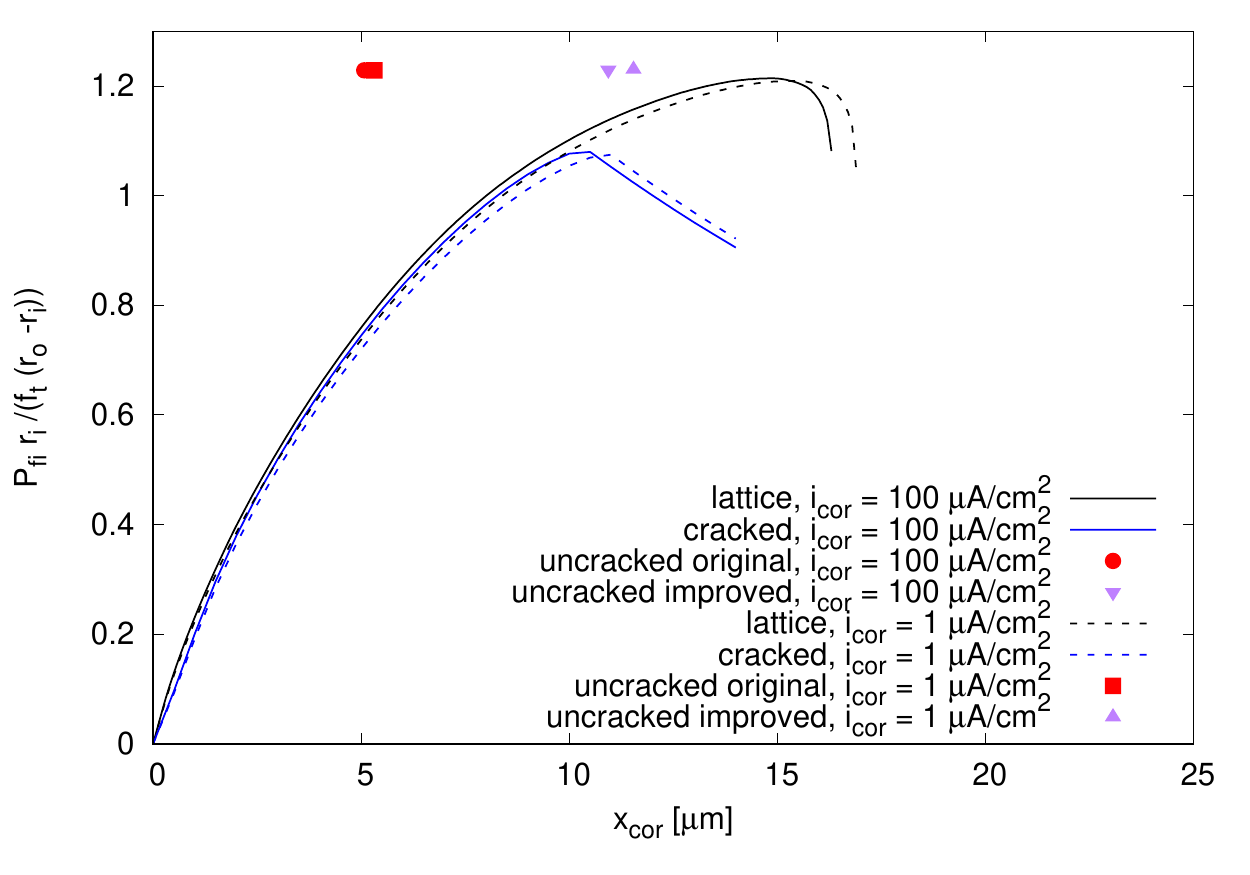}
  \end{center}
  \caption{Normalised pressure at inner boundary versus corrosion penetration $x_{\rm cor}$ for $t_0=10000$~days.}
  \label{fig:pressure10000}
\end{figure}

\begin{figure}
  \begin{center}
    \begin{tabular}{ccc}
      \includegraphics[width=5cm]{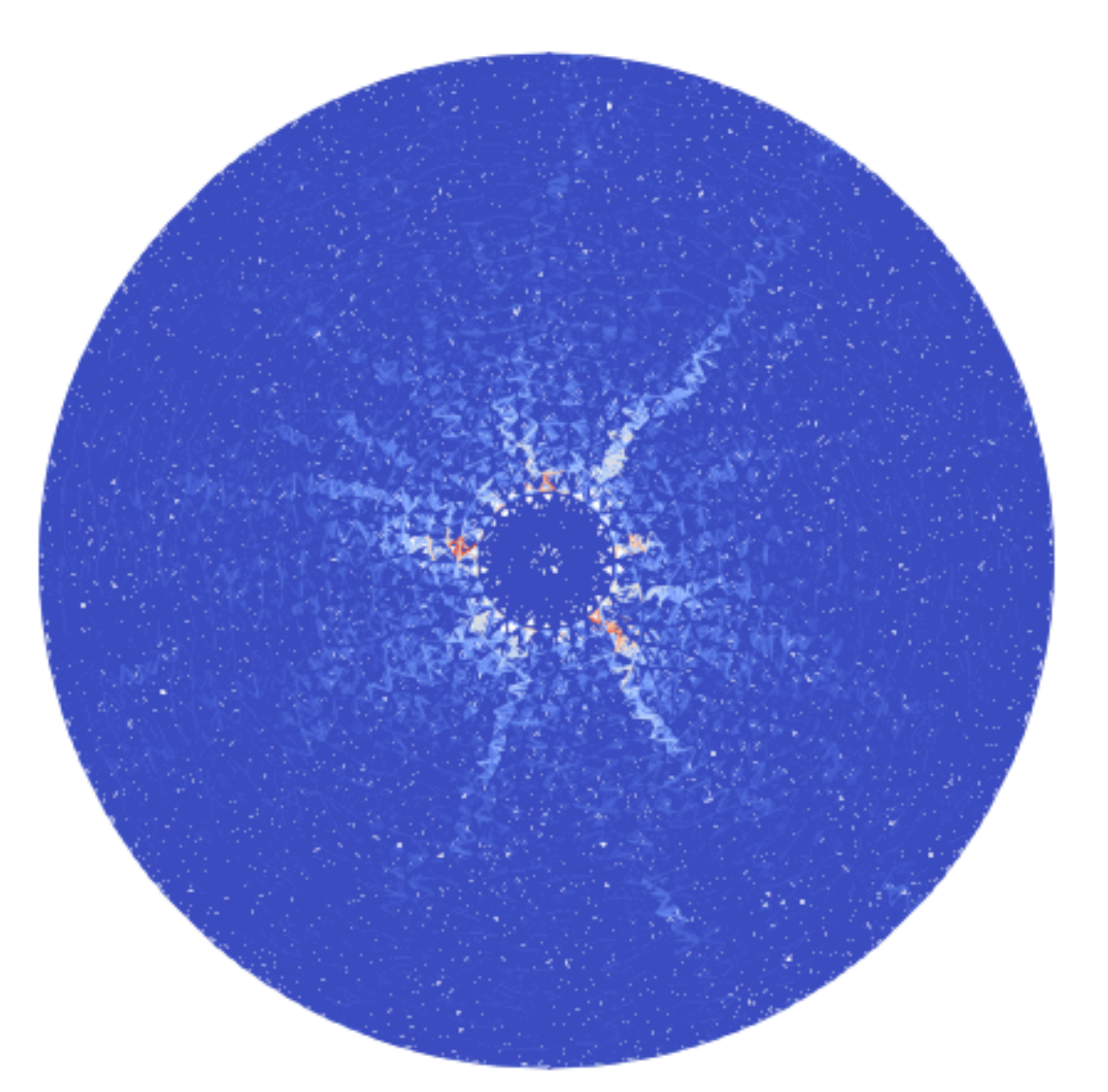} & \includegraphics[width=5cm]{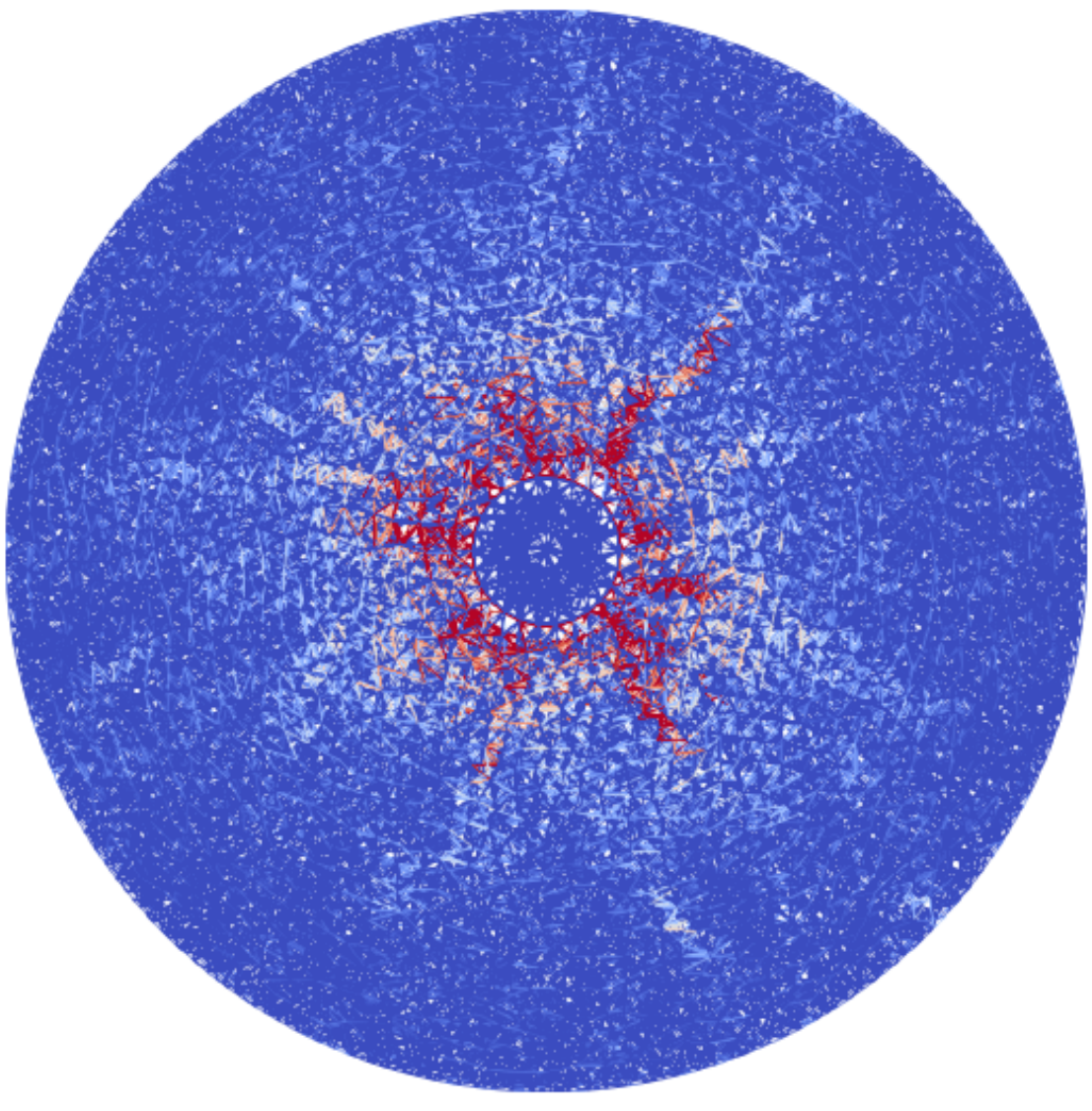} & \includegraphics[width=5cm]{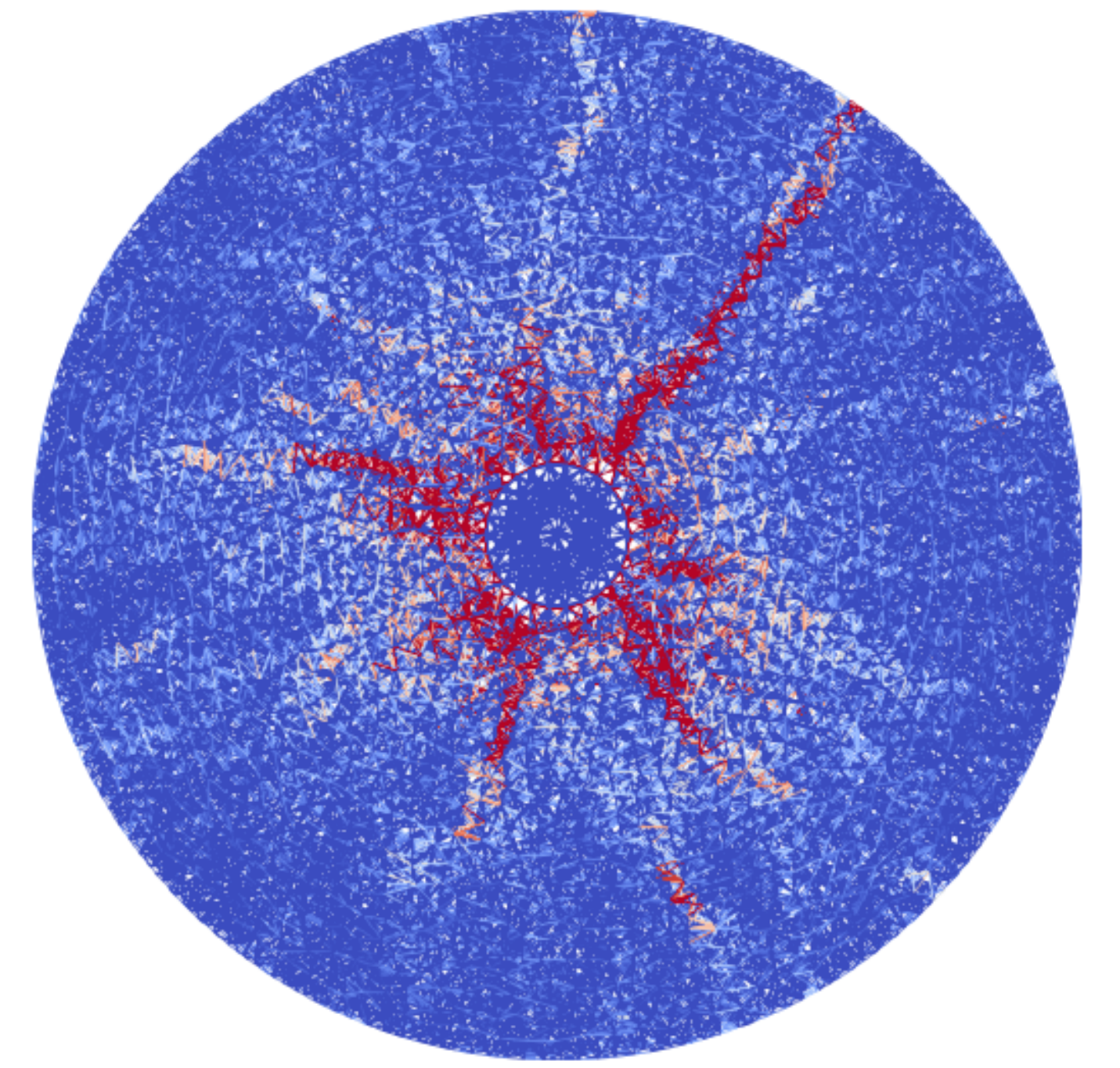}\\
      (a) & (b) & (c)\\
      \end{tabular}
  \end{center}
  \caption{Crack patterns of lattice analyses: (a)-(c) three steps marked in Figure~\ref{fig:pressure28}. Mid-cross-sections of lattice elements with crack openings greater than 2~$\mu$m are shown in red.}
  \label{fig:latticeCylinderCrack}  
\end{figure}

The main set of results of this study is the dependence of the critical corrosion penetration $x_{\mathrm{cor}}^{\mathrm{crit}}$ on the corrosion current density $i_{\mathrm cor}$. For the uncracked thick-walled cylinder model, the critical corrosion penetration $x_{\mathrm{cor}}^{\mathrm{crit}}$ is the direct output of the calculation method. For the other two models, the critical corrosion penetration is obtained from the peak of the pressure corrosion penetration curve. The peak coincides with the stage at which the deformations localise and the cracks reach the surface of the thick-walled cylinder.
This process is illustrated for $i_{\rm cor} = 100$~and~1~$\mu$A/cm$^2$ in Figures~\ref{fig:pressure28}~and~\ref{fig:pressure10000} for $t_{0} = 28$~and~$10000$~days, respectively. The result of the two versions of the uncracked cylinder model are shown as isolated points, since in this model only the critical corrosion penetration is determined. For the lattice model, the result of one of the six random analyses is shown so that the crack patterns depicted in Figure~\ref{fig:latticeCylinderCrack} can be directly linked to points in the pressure-penetration curve.  
The pressure in Figures~\ref{fig:pressure28}~and~\ref{fig:pressure10000} is normalised by $f_{\rm t}^{\rm{ref}}\left(r_{\rm o} - r_{\rm i}\right)/r_{\rm i}$, which corresponds to the pressure equilibrated by constant circumferential stress equal to the tensile strength $f_{\rm t}^{\rm{ref}}$.
The crack patterns obtained from the lattice model with $i_{\rm cor} = 100$~$\mu$A/cm$^2$ and $t_{\rm 0} = 28$~days at stages marked in Figure~\ref{fig:pressure28} are shown in Figure~\ref{fig:latticeCylinderCrack}.
These crack patterns are representative of the other lattice analyses with different $i_{\rm cor}$ values and starting age $t_0 = 10000$~days. During the first stage, cracking is rather distributed. Then, one dominant crack forms and propagates to the specimen boundary. At the peak pressure, the crack is almost connected to the boundary, which changes the deformed shape of the cylinder significantly. Based on the evolution of the crack patterns, the peak of the pressure versus corrosion penetration curve is a good criterion for surface cracking. The corrosion penetration at peak will be considered as the critical corrosion penetration $x_{\rm cor}^{\mathrm{crit}}$.

\begin{figure}
  \begin{center}
      \includegraphics[width=12cm]{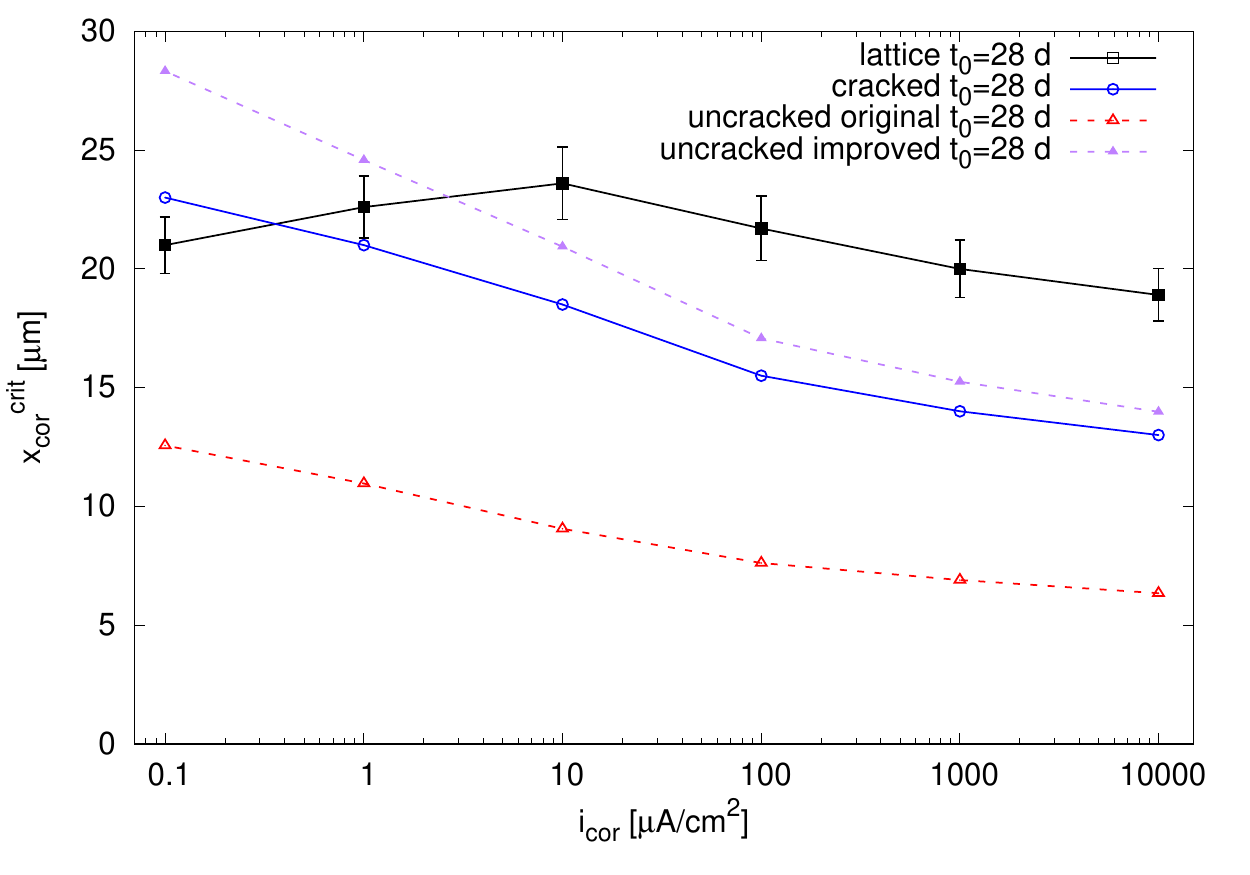}\\
  \end{center}
  \caption{
  Critical penetration depth $x_{\rm cor}^{\rm crit}$ versus corrosion rate $i_{\rm cor}$ for loading at age $t_{\rm 0} = 28$~days.}
  \label{fig:peaks28}  
\end{figure}

\begin{figure}
  \begin{center}
      \includegraphics[width=12cm]{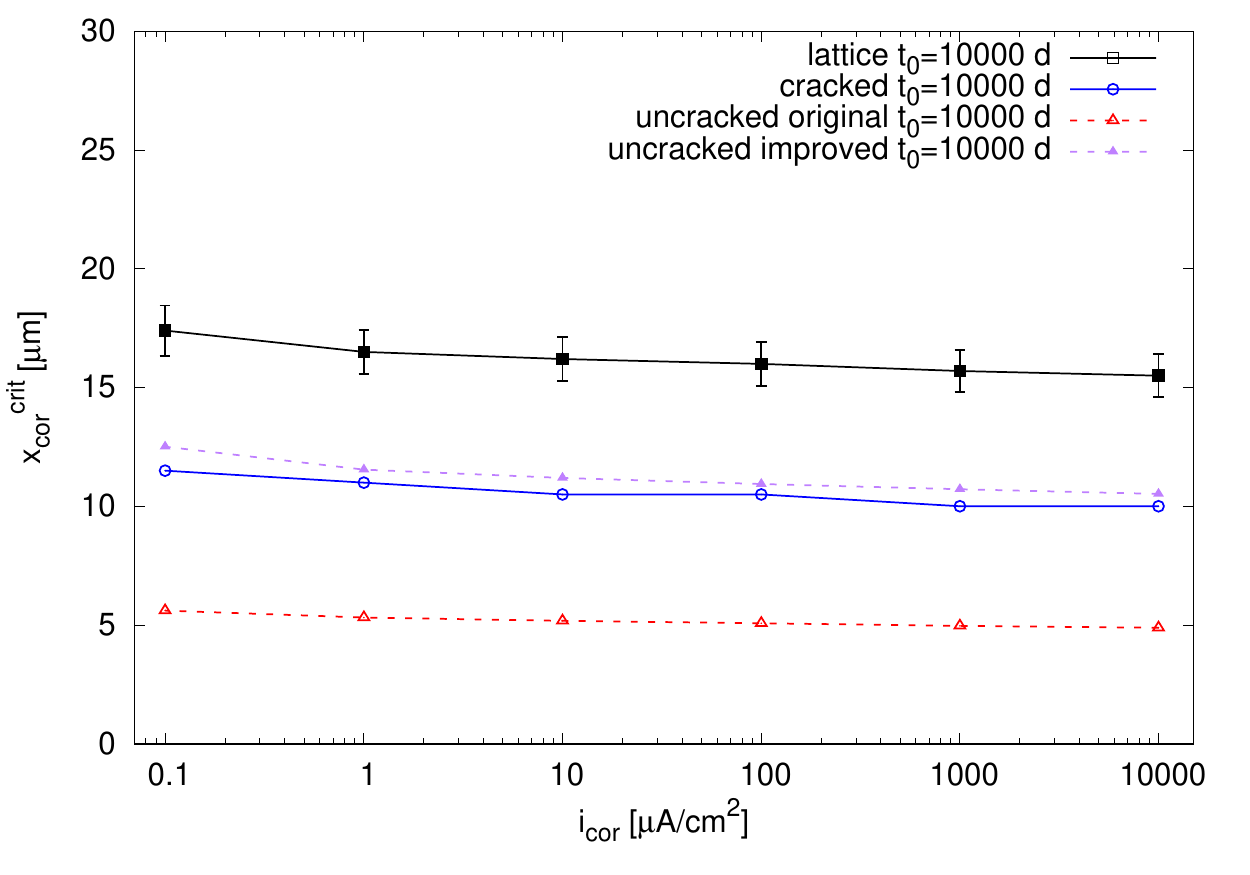}\\
  \end{center}
  \caption{
  Critical penetration depth $x_{\rm cor}^{\rm crit}$ versus corrosion rate $i_{\rm cor}$ for loading at age $t_{\rm 0} = 10000$~days.}
  \label{fig:peaks10000}  
\end{figure}

The critical corrosion penetration versus the corrosion rate is presented for the two starting ages $t_{\rm 0}=28$~and~$10000$~days in Figures~\ref{fig:peaks28}~and~\ref{fig:peaks10000} for the three modelling approaches. For the lattice model, the results are the average of six random analyses. The error bars in the figures indicate the +/- standard deviation.
The results for $t_{\rm 0}=28$~days in Figure~\ref{fig:peaks28} show that the effect of $i_{\rm cor}$ depends strongly on the modelling approach. For the uncracked model, $i_{\mathrm{cor}}$ has a very strong effect on the critical corrosion penetration. For $i_{\rm cor} = 0.1$~$\mu$A/cm$^2$,  $x_{\mathrm{cor}}^{\mathrm{crit}}$  is about twice the value for $i_{\rm cor} = 1000$~$\mu$A/cm$^2$. This strong effect is also visible for the cracked model, but not for the lattice model. With the lattice model, the critical corrosion penetration first increases and then decreases with decreasing corrosion current density. The difference between the critical corrosion penetration at $i_{\rm cor} = 10000$~$\mu$A/cm$^2$ and $i_{\rm cor} = 0.1$~$\mu$A/cm$^2$ is less than $10$~percent.
For $t_{\rm 0}=10000$~days in Figure~\ref{fig:peaks10000}, the critical corrosion penetration increases monotonically with decreasing corrosion rate for all three models. However, the increase of $x_{\mathrm{cor}}$ with increasing $i_{\mathrm{cor}}$ is much less pronounced than it was for $t_0=28$~days.

\begin{figure}
  \begin{center}
      \includegraphics[width=12cm]{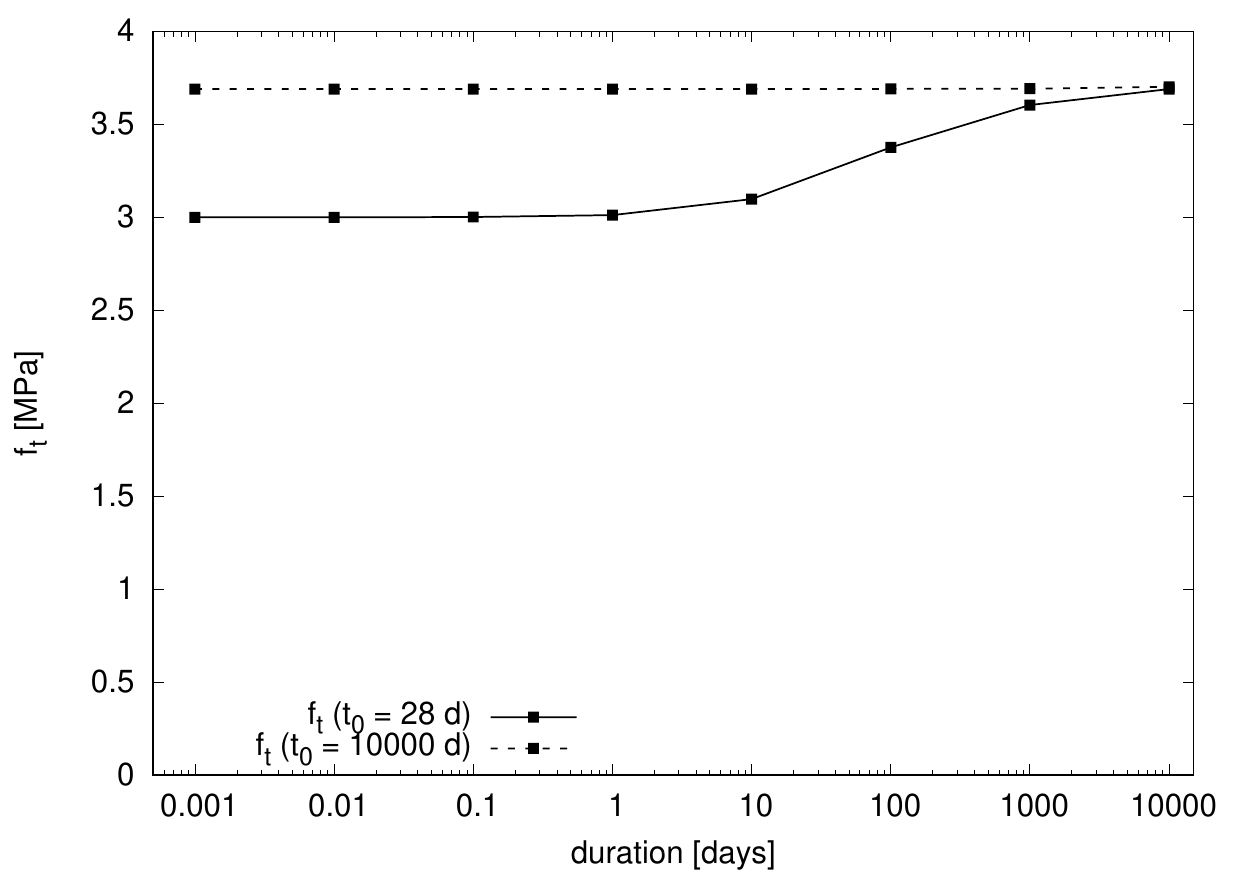}\\
  \end{center}
  \caption{Tensile strength versus loading duration for loading at $t_0=28$~and~10000~days.}
  \label{fig:crackedPeaks}  
\end{figure}

\begin{figure}
  \begin{center}
      \includegraphics[width=12cm]{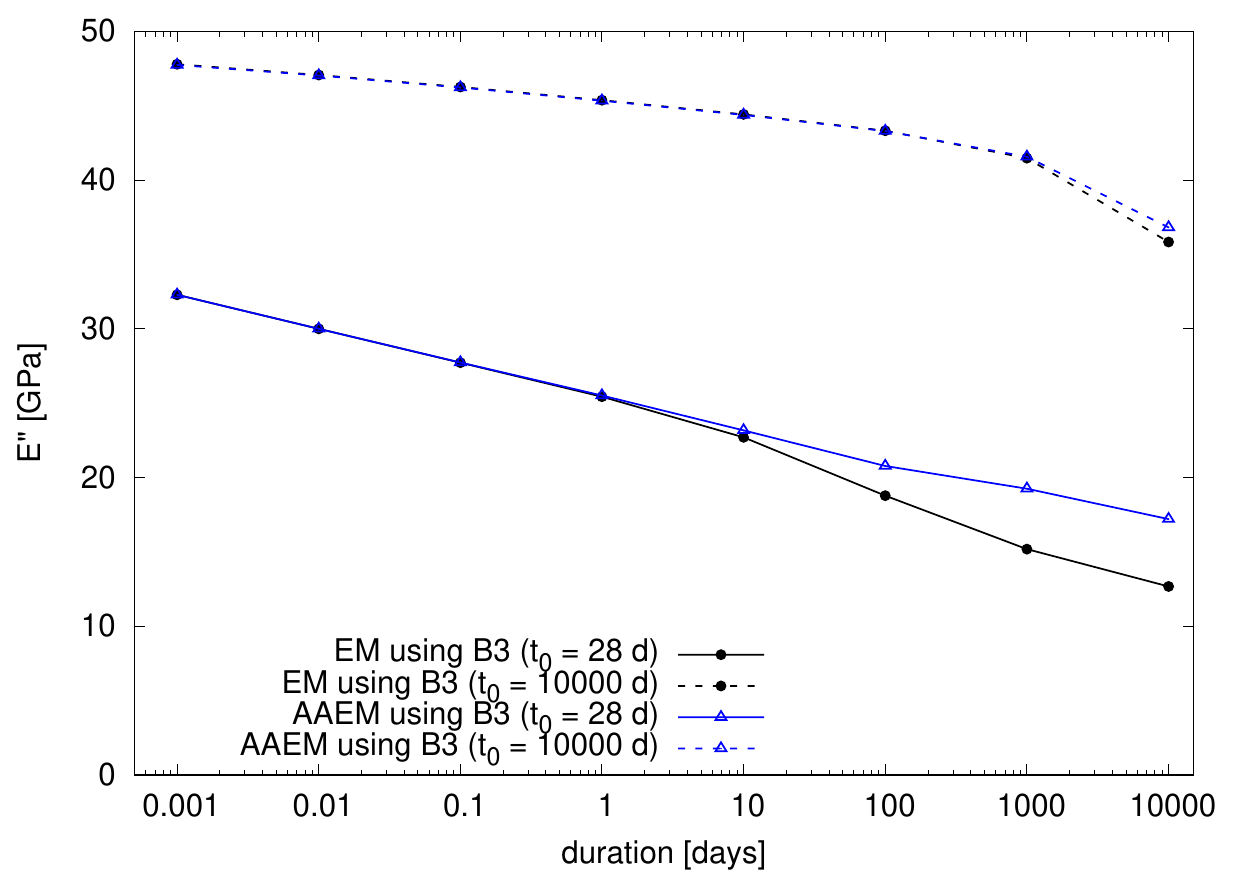}\\
  \end{center}
  \caption{Effective modulus versus loading duration for start of loading at $t_0=28$~and~10000~days.}
  \label{fig:effectiveModulus}  
\end{figure}

The trends in Figures~\ref{fig:peaks28}~and~\ref{fig:peaks10000} are the result of a number of factors, which are discussed here in more detail.
Firstly, tensile strength evolves strongly with time. As concrete matures, tensile strength increases, as shown in Figure~\ref{fig:crackedPeaks}.
The greater the tensile strength is, the more inner pressure the thick-walled cylinder can sustain, i.e., the greater is the critical corrosion penetration. The increase of tensile strength is much more pronounced for a loading at $t_0=28$~days than at $t_0=10000$~days. Therefore, for corrosion in mature concrete, changes in tensile strength are of minor importance. However, the change is important for laboratory tests for which corrosion commences soon after the concrete reaches its design strength and a small $i_{\rm cor}$ (long loading duration) is used. 
Concrete maturity at loading and load duration influence also the Young modulus of concrete.
On the one hand, Young's modulus increases with age independently of the loading regime. On the other hand, concrete subjected to load undergoes creep, which decreases the effective modulus. In Figure~\ref{fig:effectiveModulus}, the effective modulus versus the load duration is shown for  loading at $t_0=28$~and~10000~days. 
For the AAEM approach, the effective modulus is calculated from \eqref{eq:elastic8}. For the lattice model, the effective modulus is determined from a single lattice element subjected to direct tension kept constant over time. By evaluating the secant modulus at increasing times, the evolution of the effective modulus is obtained. The values of the effective moduli for the lattice model were scaled so that the effective modulus for $t_0=28$ and $\Delta t = 0.01$~days is equal to the continuum Young's modulus ${E}^{28}$, which permits a direct comparison of the effective modulus and AAEM.
For the AAEM approach used for the uncracked and cracked thick-walled cylinder model, the age-adjusted effective modulus ${E}^{''}(t_0=28,\Delta t = 0.01)$ is equal to the target reference Young's modulus ${E}^{28}$. It can be seen that the maturity of concrete strongly affects the effective modulus. The increase in the effective modulus due to aging is stronger than the decrease due to creep. Furthermore, the younger the concrete is, the more the effective modulus decreases due to creep. Comparing AAEM and EM, it is visible that the reduction of the effective modulus due to creep is stronger for the AAEM approach. This agrees with the trends presented in \citet{BazJir18}.

The greater the effective modulus is, the smaller is $x_{\mathrm{cor}}^{\mathrm{crit}}$, because a stiffer material around the expanding layer of corrosion products will produce higher stresses. This effect is the most visible for the original version of the uncracked model in Figure~\ref{fig:peaks28}, because in this model there is no reduction of stiffness due to cracking, whereas for the cracked and lattice model cracking is modelled, and the improved version of the uncracked model takes into account plastic strains. The critical corrosion penetration $x_{\mathrm{cor}}^{\mathrm{crit}}$ is the least affected by changes in the effective modulus in the lattice approach. In Figure~\ref{fig:peaks28}, it is even visible that the critical corrosion penetration value decreases for corrosion current densities less than $10$~$\mu$A/cm$^2$ in the lattice model. This reversal in the trend of critical corrosion penetration can be explained by the competing influences of the evolution of tensile strength and Young's modulus. As explained above, aging of concrete increases the tensile strength and Young's modulus. An increase of tensile strength increases the critical corrosion penetration whereas an increase in Young's modulus results in a reduction of this penetration. The way in which these aging driven changes interact depends on the modelling approach.

\section{Conclusions}
Three modelling approaches of varying complexity have been used in this study of the effects of linear creep and aging of concrete on initiation of corrosion-induced surface cracking.

The overall conclusion from this study is that the more comprehensive the model is, the less important is the effect of linear creep and aging of concrete on corrosion-induced cracking. For corrosion starting at the age of 28~days, both creep and increase in strength and stiffness due to increasing maturity are very pronounced. However, the interaction of these competing processes results in critical corrosion penetrations which are not too sensitive to the corrosion rate. For corrosion in mature concrete, both creep and change in strength and stiffness are small. For the simpler models, particularly the uncracked model, the corrosion rate has a strong effect on the critical corrosion penetration for corrosion starting at $t_0 = 28$~days.
This conclusion is important, since simpler mechanistic models use creep coefficients  which produce critical corrosion penetrations that are very sensitive to the corrosion rate, as it was shown for the simpler models in the present study as well. Our finding with the comprehensive model shows that creep is not the source of sensitivity to corrosion rate. Instead, other variables should be introduced in predictive models which can explain the sensitivity of critical corrosion penetration to corrosion rate. Candidates for these variables would be the viscosity and expansion coefficient of corrosion products, as well as penetration of corrosion products into pores and cracks.

Other conclusions are that it has been shown that the three models predict the critical corrosion penetration very differently for fast corrosion rates, for which linear creep and aging have a small effect on the results. The uncracked model, which is the simplest of the three approaches, predicts in its original version the smallest critical corrosion penetration because it does not take into account the reduction of stiffness of the concrete due to cracking. The cracked model predicts greater critical corrosion penetrations because it models the cracking process. The improved version of the uncracked model accounts for plastic strains and provides an analytical upper bound to the predictions obtained with the cracked model.
Finally, the lattice modelling approach usually predicts the greatest critical corrosion penetration
(with the exception of very low corrosion rates applied at a young age) because it models also distributed cracking and nonlinearities in compression, which is not considered in the cracked model. These results are in agreement with the overall findings in the literature. The lattice model is, among the three approaches examined here, the only one which is general enough to consider nonuniform corrosion and more complex stress states.

\section*{Acknowledgements}

Aldellaa is grateful for the financial support from the Ministry of Higher Education - Libya.
Havl\'{a}sek acknowledges the support of the Czech Science Foundation [grant GA~\v{C}R~19-20666S].
Jir\'{a}sek acknowledges the support of the European Regional Development Fund 
(Center of Advanced Applied Sciences, project CZ.02.1.01/\-0.0/\-0.0/\-16\_19/\-0000778).
The lattice simulations were carried out with the finite element programme OOFEM \citep{Pat12} modified by the authors.

\bibliographystyle{kbib}
\bibliography{general}

\end{document}